\documentclass{elsart}

\usepackage{natbib}
\usepackage{graphicx}
\usepackage{amssymb}

\newcommand{\cc}{\mbox{\boldmath$c$}}
\newcommand{\ee}{\mbox{\boldmath$e$}}
\newcommand{\ff}{\mbox{\boldmath$f$}}
\newcommand{\hh}{\mbox{\boldmath$h$}}
\newcommand{\bl}{\mbox{\boldmath$l$}}
\newcommand{\yy}{\mbox{\boldmath$y$}}
\newcommand{\xxx}{\mbox{\boldmath$x$}}
\newcommand{\xx}{\mbox{\boldmath$x$}}
\newcommand{\zz}{\mbox{\boldmath$z$}}
\newcommand{\bb}{\mbox{\boldmath$b$}}
\newcommand{\bg}{\mbox{\boldmath$g$}}
\newcommand{\mm}{\mbox{\boldmath$m$}}
\newcommand{\vl}{\mbox{\boldmath$l$}}
\newcommand{\BB}{\mbox{\boldmath$B$}}
\newcommand{\CC}{\mbox{\boldmath$C$}}
\newcommand{\DD}{\mbox{\boldmath$D$}}
\newcommand{\FF}{\mbox{\boldmath$F$}}
\newcommand{\VV}{\mbox{\boldmath$V$}}
\newcommand{\JJ}{\mbox{\boldmath$J$}}
\newcommand{\PP}{\mbox{\boldmath$P$}}
\newcommand{\HH}{\mbox{\boldmath$H$}}
\newcommand{\LL}{\mbox{\boldmath$L$}}

\newcommand{\NN}{\mbox{\boldmath$N$}}
\newcommand{\UU}{\mbox{\boldmath$U$}}
\newcommand{\ZZ}{\mbox{\boldmath$Z$}}

\newcommand{\bZERO}{\mbox{\boldmath$0$}}
\newcommand{\bOmega}{\mbox{\boldmath$\Omega$}}
\newcommand{\bPi}{\mbox{\boldmath$\Pi$}}
\newcommand{\bPsi}{\mbox{\boldmath$\Psi$}}
\newcommand{\bDelta}{\mbox{\boldmath$\Delta$}}
\newcommand{\bgamma}{\mbox{\boldmath$\gamma$}}
\newcommand{\bmu}{\mbox{\boldmath$\mu$}}
\newcommand{\bnabla}{\mbox{\boldmath$\nabla$}}
\newcommand{\balpha}{\mbox{\boldmath$\alpha$}}
\newcommand{\bbeta}{\mbox{\boldmath$\beta$}}

\begin{document}

\begin{frontmatter}

\title{Method of invariant manifold \\for chemical kinetics}

\author{Alexander N.\ Gorban}
\address{Department of Materials, Institute of Polymer Physics\\
Swiss Federal Institute of Technology, CH-8092 Z\"urich, Switzerland
\\Institute of Computational Modeling RAS, Krasnoyarsk 660036, Russia}

\author{Iliya V.\ Karlin}
\address{Department of Materials, Institute of Polymer Physics\\
Swiss Federal Institute of Technology, CH-8092 Z\"urich, Switzerland}


\begin{abstract}
In this paper, we review the construction of low-dimensional manifolds of reduced description
for equations of chemical kinetics from the standpoint of the method of invariant 
manifold (MIM). MIM is based on a formulation of the condition of invariance as
an equation, and its solution by Newton iterations. 
A review of existing alternative methods is extended by
a thermodynamically consistent version of the method of  intrinsic low-dimensional manifolds.
A grid-based version of MIM is developed, and model extensions of low-dimensional
dynamics are described. Generalizations to open systems are suggested.
The set of methods covered makes it possible to effectively reduce description
in chemical kinetics.

\end{abstract}


\end{frontmatter}
\section{Introduction}
In this paper, we present  a general method of
constructing the reduced description for dissipative systems of reaction
kinetics. Our approach is based on the method of invariant manifold which
was developed  in end of 1980th - beginning of 1990th by
\cite{GK92,GK92a,GK92b}.
Its realization for a generic
dissipative systems was discussed by \cite{GK94,GKIOe01}.
This
method was applied to a set of  problems of classical kinetic
theory based on the Boltzmann kinetic equation
\citep{GK94,KDN97,KGDN98}.
The method of invariant manifold was  successfully applied to a derivation of
reduced description for kinetic equations of polymeric solutions
\citep{ZKD00}. 
It was also been tested on systems of chemical kinetics \citep{GKZD00}.

The goal of nonequilibrium statistical physics is the
understanding of how a system with many degrees of freedom
acquires a description with a few degrees of freedom.
This should lead to reliable methods of extracting
the macroscopic description from a detailed microscopic description.

Meanwhile this general problem is still far from the final solution, it is
reasonable to study simplified models, where, on the one hand, a
detailed description is accessible to numerics, on the other hand,
analytical methods designed to the solution of problems in real systems
can be tested.

In this paper we address the well known class of finite-dimensional
systems known from the theory of reaction kinetics. 
These are equations governing a complex relaxation in perfectly stirred closed
chemically active mixtures.
Dissipative properties of such systems are characterized with a global
convex Lyapunov function $G$ (thermodynamic potential) which implements the second law of
thermodynamics: As the time $t$ tends to infinity, the system reaches the
unique equilibrium state while in the course of the transition the Lyapunov
function decreases monotonically.

While the limiting behavior of the dissipative systems just described
is certainly very simple, there are still interesting questions to be
asked about. One of these questions is closely related to the above
general problem of nonequilibrium statistical physics. Indeed,
evidence of numerical integration of such systems
often demonstrates that the relaxation
has a certain geometrical structure in the phase space.
Namely, typical individual trajectories
tend to manifolds of lower dimension, and further proceed to
the equilibrium essentially along these manifolds.
Thus, such systems demonstrate a dimensional reduction, and therefore
establish a more macroscopic description after some time since
the beginning of the relaxation.

There are two intuitive ideas behind our approach, and we shall now
discuss them informally. Objects to be considered below are manifolds
(surfaces) $\bOmega$ in the phase space 
of the reaction kinetic system
(the phase space is usually a convex polytope in a finite-dimensional real space).
The `ideal' picture of the reduced description we have in mind is as
follows: A typical phase trajectory, ${\cc}(t)$,
where $t$ is the time, and $\cc$ is an element of the phase space,
consists of two pronounced segments. The first segment connects the
beginning of the trajectory, $\cc(0)$, with a certain point, $\cc(t_1)$,
on the manifold $\Omega$ (rigorously speaking, we should think
of  $\cc(t_1)$ not on $\bOmega$ but in a small neighborhood of $\bOmega$
but this is inessential for the ideal picture). The second segment belongs  to $\bOmega$,
and connects the point $\cc(t_1)$ with the equilibrium
$\cc^{\rm eq}={\cc}(\infty)$, $\cc^{\rm eq}\in\bOmega$.
Thus, the manifolds appearing in our ideal picture are
``patterns'' formed by the segments of individual trajectories,
and the goal of the reduced description is to ``filter out'' this manifold.

There are two important features behind this ideal picture. The first feature 
is the {\it invariance} of the manifold $\bOmega$: Once the individual
trajectory has started on  $\bOmega$, it does not leaves $\bOmega$
anymore. The second feature is the {\it projecting}: The phase points 
outside $\bOmega$ will be projected onto $\bOmega$. 
Furthermore, the dissipativity of the system provides an
{\it additional} information about this ideal picture:
Regardless of what happens on  the manifold $\bOmega$,
the function $G$ was decreasing along each individual trajectory
before it reached $\bOmega$. This ideal picture is the guide to extract slow invariant
manifolds.

The paper is organized as follows. In the section \ref{OUTLINE}, we review
the reaction kinetics (section \ref{KINETICS}), 
and discuss the main methods of model reduction in chemical kinetics 
(section \ref{reduction_review}).
In particular, we present two general
versions of extending partially equilibrium manifolds
to a single relaxation time model in the whole
phase space, and develop a thermodynamically consistent version
of the intrinsic low-dimensional manifold (ILDM) approach.
In the section \ref{MIMG} we  introduce the method of invariant manifold
in the way appropriate to this class of nonequilibrium systems.
In the sections \ref{THERMO} and \ref{DC} we give some
details on the two relatively independent parts of the method,
the thermodynamic projector, and the iterations for solving the invariance
equation. We also introduce a general
symmetric linearization procedure
for the invariance equation, and discuss its relevance to the picture
of decomposition of motions. In the section \ref{MIM}, these two
procedures are combined into an unique algorithm. In the section \ref{EX},
we demonstrate an illustrative example of computations for a model catalytic
reaction. 
In the section \ref{PARAMETERIZATION}
we demonstrate how the thermodynamic 
projector is constructed without the a priori parameterization of the manifold.
This result is essentially used 
in the section \ref{grid} where we introduce  
a computationally effective grid-based method to construct  invariant manifolds.
In the section \ref{open} we describe an extension of the method
of invariant manifold to open systems.
Finally, results are discussed in the section \ref{conclusion}.

\section{Equations of chemical kinetics and their reduction} 
\label{OUTLINE}
\subsection{Outline of the dissipative reaction kinetics}
\label{KINETICS}
We begin with an outline  of the
reaction kinetics (for details see e.\ g.\ the book of \cite{YBGE91}).
Let us consider a closed system
with $n$ chemical species ${\rm A}_1,\dots,{\rm A}_n$,
participating in a complex reaction. The complex reaction is
represented by the following stoichiometric mechanism:
\begin{equation}
\label{stoi}
\alpha_{s1}{\rm A}_1+\ldots+\alpha_{sn}{\rm A}_n\rightleftharpoons
\beta_{s1}{\rm A}_1+\ldots+\beta_{sn}{\rm A}_n,
\end{equation}
where the index $s=1,\dots,r$ enumerates the reaction steps, and
where integers, $\alpha_{si}$ and $\beta_{si}$,  are stoichiometric
coefficients. For each reaction step $s$, we introduce 
$n$--component vectors $\balpha_s$ and $\bbeta_s$ with components
$\alpha_{si}$ and $\beta_{si}$. Notation
$\mbox{\boldmath$\gamma$}_s$  stands for the vector
with integer components
$\gamma_{si}=\beta_{si}-\alpha_{si}$ (the stoichiometric vector). 
We adopt an abbreviated notation for the
standard scalar product of the $n$-component vectors:
\[\langle\xx,\yy\rangle=\sum_{{i=1}}^{n}x_iy_i.\]

The system is described by the  $n$-component concentration vector
$\cc$, where the component $c_i\ge0$ represents the
concentration of the specie ${\rm A}_i$. Conservation laws impose linear
constraints on admissible vectors $\cc$ (balances):
\begin{equation}
\label{conser}
\langle\bb_i, \cc\rangle=B_i,\ i=1,\dots,l,
\end{equation}
where $\bb_i$ are fixed and linearly independent vectors, 
and $B_i$ are given scalars. Let us denote as $\BB$ the set of vectors
which satisfy the conservation laws (\ref{conser}):
\[
\BB=\left\{\cc|\langle\bb_1,\cc\rangle=B_1,\dots,
\langle\bb_l,\cc\rangle=B_l\right\}.
\]
The phase space $\VV$ of the system is the intersection 
of the cone of $n$-dimensional vectors with nonnegative components,
with the set $\BB$, and ${\rm dim}\VV=d=n-l$. 
In the sequel, we term a vector $\cc\in\VV$ the state
of the system. In addition, we assume that each of the conservation
laws is supported by each elementary reaction step, that is
\begin{equation}
\label{sep}
\langle\bgamma_s,\bb_i\rangle=0,
\end{equation}
for each pair of vectors $\bgamma_s$ and $\bb_i$.

Reaction kinetic equations describe variations of the states in time.
Given the stoichiometric mechanism (\ref{stoi}),
the reaction kinetic equations read:
\begin{equation}
\label{reaction}
\dot{\cc}=\JJ(\cc),\ 
\JJ(\cc)=\sum_{s=1}^{r}\bgamma_sW_s(\cc),
\end{equation}
where dot denotes the time derivative, and $W_s$  is the 
reaction rate function of the step $s$.
In particular, the mass action law suggests the polynomial form of the
reaction rates:
\begin{equation}
\label{MAL}
W_s=k^+_s \prod_{i=1}^{n}c_i^{\alpha_i} -
k^-_s \prod_{i=1}^{n}
c_i^{\beta_i},
\end{equation}
where $k^+_s$ and $k^-_s$ are the constants of the direct and of
the inverse reactions rates of the $s$th reaction step.
The phase space $\VV$ is positive-invariant of the system
(\ref{reaction}): If $\cc(0)\in\VV$, then $\cc(t)\in\VV$
for all the times $t>0$.

In the sequel, we assume that the kinetic equation (\ref{reaction})
describes  evolution towards the unique equilibrium state, $\cc^{\rm eq}$,
in the interior of the phase space $\VV$. Furthermore, we assume
that there exists a strictly convex function
$G(\cc)$ which decreases monotonically in time due to Eq.\
(\ref{reaction}):
\begin{equation}
\label{Htheorem}
\dot{G}=
\langle\bnabla G(\cc),\JJ(\cc)\rangle
\leq 0,
\end{equation}
Here $\bnabla G$ is the
vector of partial derivatives $\partial G/\partial c_i$, and the convexity
assumes that the $n\times n$ matrices
\begin{equation}
\label{MATRIX}
\HH_{\cc}=\|\partial^2G(\cc)/\partial c_i\partial c_j\|,
\end{equation}
are positive definite for all $\cc\in\VV$. In addition, we assume that the
matrices (\ref{MATRIX}) are invertible if $\cc$ is taken in
the interior of the phase space.

The function $G$ is the Lyapunov function of the system
(\ref{reaction}), and $\cc^{\rm eq}$ is the point of global minimum of
the function $G$ in the phase space $\VV$. Otherwise stated,
the manifold of equilibrium states $\cc^{\rm eq}(B_1,\dots,B_l)$
is the solution to the variational problem,
\begin{equation}
\label{EQUILIBRIUM}
G\to{\rm min}\ {\rm for\ }\langle\bb_i,\cc\rangle=B_i,\ i=1,\dots,l.
\end{equation}
For each fixed value of the conserved quantities $B_i$, the solution
is unique. In many cases, however, it is convenient to consider
the whole equilibrium manifold, keeping the conserved quantities
as parameters.

For example, for perfect systems in a constant volume under a constant
temperature, the Lyapunov function $G$ reads:
\begin{equation}
\label{gfun}
G=\sum_{i=1}^{n}c_i[\ln(c_i/c^{\rm eq}_i)-1].
\end{equation}

It is important to stress that $\cc^{\rm eq}$ in Eq.\ (\ref{gfun})
 is an {\it arbitrary} equilibrium of the system,
under arbitrary values of the balances. In order to compute $G(\cc)$, 
it is unnecessary to calculate
the specific equilibrium $\cc^{\rm eq}$ which corresponds to the initial state $\cc$.
Moreover, for ideal systems, function $G$ is constructed from the thermodynamic 
data of individual
species, and, as  the result of this construction, it turns out that it has the form of 
Eq.\ (\ref{gfun}).
Let us mention here the classical formula for the free energy $F=RTVG$: 
\begin{equation}
F=VRT\sum_{i=1}^{n}c_i[(\ln(c_iV_{{\rm Q}\ i})-1)+F_{{\rm int}\ i}(T)],
\end{equation}
where $V$ is the volume of the system, $T$ is the temperature, 
$V_{{\rm Q}\ i}= N_0(2\pi\hbar^2/m_ikT)^{3/2}$
is the quantum volume of one mole of the specie $A_i$, $N_0$ is the Avogadro number,
$m_i$ is the mass of the molecule of ${\rm A}_i$, $R=kN_0$, and $F_{{\rm int}\ i}(T)$ is the
free energy of the internal degrees of freedom per mole of ${\rm A}_i$.

Finally, we recall an important generalization of the mass action
law (\ref{MAL}), known as the Marcelin-De Donder kinetic function.
This generalization was developed by \cite{Feinberg72} based on 
ideas of the thermodynamic theory of affinity \citep{DeDonder36}. 
We use
the kinetic function suggested by \cite{BGY82}.
Within this approach, the functions $W_s$ are
constructed as follows: For a given strictly convex function $G$,
and for a given stoichiometric mechanism (\ref{stoi}), we define the gain
($+$) and the loss ($-$) rates of the $s$th step,
\begin{equation}
\label{MDD}
W_s^{+}=\varphi_s^+
\exp[\langle \nabla G,\mbox{\boldmath$\alpha$}_s\rangle],\quad
W_s^{-}=\varphi_s^-\exp[\langle \nabla G,\mbox{\boldmath$\beta$}_s\rangle],
\end{equation}
where $\varphi_s^{\pm}>0$ are kinetic factors.
The Marcelin-De Donder kinetic function reads: $W_s=W_s^+-W_s^-$, 
and the right hand side of the kinetic equation (\ref{reaction}) becomes,
\begin{equation}
\label{KINETIC MDD}
\JJ=\sum_{s=1}^{r}\bgamma_s
\{\varphi_s^+
\exp[\langle\bnabla G,\balpha_s\rangle]\!-
\varphi_s^-\exp[\langle\bnabla G,\bbeta_s\rangle]\}.
\end{equation}
For the Marcelin-De Donder reaction rate (\ref{MDD}), the dissipation
inequality (\ref{Htheorem}) reads:
\begin{equation}
\label{HMDD}
\dot{G}=\sum_{s=1}^{r}
[\langle\bnabla G,\bbeta_s\rangle -
\langle\bnabla G,\balpha_s\rangle]
\left\{\varphi_s^+
e^{\langle\bnabla G,\balpha_s\rangle}-
\varphi_s^-e^{\langle\bnabla G,\bbeta_s\rangle}\right\}\le 0.
\end{equation}
The kinetic factors $\varphi_s^{\pm}$ should satisfy certain conditions
in order to make valid the dissipation inequality (\ref{HMDD}).
A well known sufficient condition is the detail balance:
\begin{equation}
\label{DB}
\varphi_s^+=\varphi_s^-,
\end{equation}
other sufficient conditions are discussed in detail elsewhere
\citep{YBGE91,G84,K89,K93}.
For the function  $G$ of the form (\ref{gfun}),
the Marcelin-De Donder equation casts into the more familiar mass action
law form (\ref{MAL}).

\subsection{The problem of reduced description in chemical kinetics}
\label{reduction_review}
What does it mean, ``to reduce the description of a chemical system''? 
This means the following:
\begin{enumerate}
\item To shorten  the list of species.
This, in turn, can be achieved in two ways: 

(i) To eliminate inessential components from the list;

(ii) To lump some of the species into integrated components.

\item To shorten the list of reactions. This also can be done in several ways:

(i) To eliminate inessential reactions, those which do not significantly influence the reaction
process;

(ii) To assume that some of the reactions ``have been already completed'', and that
the equilibrium has been reached along their paths (this leads to dimensional reduction
because the rate constants of the ``completed'' reactions are not used thereafter, what one
needs are equilibrium constants only). 

\item To decompose the motions into fast and slow, into independent (almost-independent) 
and slaved etc.
As the result of such a decomposition, the system admits a study ``in parts''. 
After that, results of this study
are combined into a joint picture. There are several approaches which fall into this category:
The famous method of the quasi-steady state (QSS), pioneered by 
Bodenstein and Semenov and 
explored in considerable detail  by many authors, in particular, by 
\cite{Bowen63,Chen88,Segel89,Fraser88,Roussel90}, 
and many others;
the quasi-equilibrium approximation \citep{Orlov84,G84,Volpert85,Fraser88,K89,K93};
methods of sensitivity analysis \citep{Rabitz83,Lam94};
methods based on the derivation of the so-called intrinsic low-dimensional manifolds (ILDM,
as suggested by \cite{Maas92}). Our method of invariant manifold 
(MIM, \citep{GK92,GK92a,GK92b,GK94,GKZD00,GKIOe01}) also belongs to this kind of methods.

\end{enumerate}
Why to reduce description in the times of supercomputers? 

First, in order to gain understanding. In the process of reducing the description one is often able
to extract the essential, and the mechanisms of the processes under study become
more transparent. Second, if one is given the detailed description of the system, then one
should be able also to solve the initial-value problem for this system. But what should one do 
in the case where the the system is representing just a point in a three-dimensional flow? 
The problem of reduction becomes particularly important for modeling the spatially distributed
physical and chemical processes. Third, without reducing the kinetic model, it is impossible to
construct this model. This statement seems paradoxal only at the first glance: How come,
the model is first simplified, and is constructed only after the simplification is done?
However, in practice, the typical for a mathematician statement of the problem,
(Let the system of differential equations be {\it given}, then ...) is rather rarely applicable
in the chemical engineering science for detailed kinetics. Some reactions are known
precisely, some other - only hypothetically. Some intermediate species are well studied,
some others - not, it is not known much about them. Situation is even worse with the
reaction rates. Quite on the contrary, the thermodynamic data 
(energies, enthalpies, entropies, chemical
potentials etc) for sufficiently rarefied systems are quite reliable. Final identification
of the model is always done on the basis of comparison with the experiment and with a help
of fitting. For this purpose, it is extremely important to reduce the dimension of the system,
and to reduce the number of tunable parameters. The normal logics of modeling for
the purpose of chemical engineering science is the following: Exceedingly detailed but coarse 
with respect to parameters system $\to$ reduction $\to$ fitting $\to$ reduced model with
specified parameters (cycles  are allowed in this scheme, with returns 
from fitting to more detailed models etc).
A  more radical viewpoint is also possible: 
In the chemical engineering science, detailed kinetics is impossible,
useless, and it does not exist. For a recently published discussion on this topic see
\cite{Levenspiel99,Levenspiel00}; \cite{Y00}.

Alas, with a mathematical statement of the problem
 related to reduction, we  all have to begin with the usual:
Let the system of differential equations be given ... . Enormous difficulties related to the question
of how well the original system is modeling the real kinetics remain out of focus of these studies.

Our present work is devoted to studying reductions in a given system of kinetic equations to
invariant manifolds of slow motions. We begin with a brief discussion of existing approaches.

\subsection{Partial equilibrium approximations}\label{partial_eq}

{\it Quasi-equilibrium with respect to reactions} is constructed as follows: 
From the list of reactions (\ref{stoi}), one selects those which are assumed to equilibrate first.
Let they be  indexed with the numbers $s_1,\dots,s_k$. 
The quasi-equilibrium manifold is defined
by the system of equations,
\begin{equation}
\label{st1}
W^+_{s_i}=W^-_{s_i},\ i=1,\dots,k.
\end{equation}
This system of equations looks particularly elegant when written in terms of conjugated (dual)
variables,  $\bmu=\bnabla G$:
\begin{equation}
\label{st2}
\langle \bgamma_{s_i},\bmu\rangle=0,\ i=1,\dots,k.
\end{equation}
In terms of conjugated variables, 
the quasi-equilibrium manifold
forms a linear subspace. This subspace, $L^{\perp}$, is the orthogonal 
completement to the linear
envelope of vectors, $L={\rm lin}\{\bgamma_{s_1},\dots,\bgamma_{s_k}\}$.

{\it Quasi-equilibrium with respect to species} is constructed practically  in the same  
way but without 
selecting the subset of reactions. For a given set of species, $A_{i_1}, \dots, A_{i_k}$, 
one assumes 
that they evolve fast to equilibrium, and remain there. Formally, this means that in the
$k$-dimensional subspace of the space of concentrations with the coordinates
$c_{i_1},\dots,c_{i_k}$, one constructs the subspace $L$ which is defined by the balance
equations, $\langle \bb_i,\cc\rangle=0$. In terms of the conjugated variables,
the quasi-equilibrium manifold, $L^{\perp}$, is defined by equations,
\begin{equation}
\label{qe1}
\bmu\in L^{\perp},\ (\bmu=(\mu_1,\dots,\mu_n)).
\end{equation}
The same quasi-equilibrium manifold can be also defined with the help of fictitious reactions:
Let $\bg_1,\dots,\bg_q$ be a basis in $L$. Then Eq.\ (\ref{qe1}) may be rewritten as follows:
\begin{equation}
\label{qe2}
\langle \bg_i,\bmu\rangle=0,\ i=1,\dots,q.
\end{equation}

{\it Illustration:} Quasi-equilibrium with respect to reactions in hydrogen oxidation:
Let us assume equilibrium with respect to dissociation reactions,
${\rm H}_2\rightleftharpoons 2{\rm H}$, and, ${\rm O}_2\rightleftharpoons 2{\rm O}$, 
in some subdomain of
reaction conditions. This gives:
\[k_1^+c_{{\rm H}_2}=k_1^-c^2_{\rm H},\ k_2^+c_{{\rm O}_2}=k_2^-c_{\rm O}^2.\]
Quasi-equilibrium with respect to species: For the same reaction, let us assume
equilibrium over ${\rm H}$, ${\rm O}$, ${\rm OH}$, and ${\rm H}_2{\rm O}_2$, 
in a subgomain of reaction conditions.
Subspace $L$ is defined by balance constraints:
\[ c_{\rm H}+c_{\rm OH}+2c_{{\rm H}_2{\rm O}_2}=0,\ c_{\rm O}+c_{\rm OH}
+2c_{{\rm H}_2{\rm O}_2}=0.\]
Subspace $L$ is two-dimensional. Its basis, $\{\bg_1,\bg_2\}$ in the coordinates
$c_{\rm H}$, $c_{\rm O}$, $c_{\rm OH}$, and $c_{{\rm H}_2{\rm O}_2}$ reads:
\[
\bg_1=(1,1,-1,0),\quad
\bg_2=(2,2,0,-1).
\]
Corresponding Eq.\ (\ref{qe2}) is:
\[ \mu_{\rm H}+\mu_{\rm O}=\mu_{\rm OH},\ 2\mu_{\rm H}+2\mu_{\rm O}=
\mu_{{\rm H}_2{\rm O}_2}.\]

{\it General construction of the quasi-equilibrium manifold}: 
In the space of concentration, one defines a subspace $L$ which satisfies 
the balance constraints:
\[ \langle \bb_i,L\rangle\equiv0.\]
The orthogonal complement of $L$ in the space with coordinates 
$\bmu=\bnabla G$ defines then the quasi-equilibrium manifold $\bOmega_{L}$.
For the actual computations, one requires the inversion from $\bmu$ to $\cc$.
Duality structure $\bmu\leftrightarrow\cc$ is well studied by many authors 
\citep{Orlov84,DKN97}. 

{\it Quasi-equilibrium projector.} It is not sufficient to just derive the manifold,
it is also required to define a {\it projector} which would transform
the vector field defined on the space of concentrations to a vector field on the manifold.
Quasi-equilibrium manifold consists of points which minimize $G$ on the affine
spaces of the form $\cc+L$. These affine  planes are hypothetic planes of fast motions
($G$ is decreasing in the course of the fast motions). Therefore, the quasi-equilibrium 
projector maps the whole space of concentrations on $\bOmega_L$ parallel to $L$.
The vector field is also projected onto the tangent space of $\bOmega_L$ parallel to
$L$.

Thus, the quasi-equilibrium approximation implies the decomposition of motions
into the fast - parallel to $L$, and the slow - along the quasi-equilibrium manifold.
In order to construct the quasi-equilibrium approximation, knowledge of reaction rate constants
of ``fast'' reactions is not required (stoichiometric vectors of all these fast reaction are
in $L$, 
$\bgamma_{{\rm fast}}\in L$, thus, knowledge of $L$ suffices), one only needs some
confidence in that they all are sufficiently fast \citep{Volpert85}.
The quasi-equilibrium manifold itself is constructed based on the knowledge of
$L$ and of $G$. Dynamics on the quasi-equilibrium manifold is defined as the quasi-equilibrium projection of the ``slow component'' of kinetic equations (\ref{reaction}).

\subsection{Model equations}

The assumption behind  the quasi-equilibrium is the hypothesis of the decomposition of 
motions into fast and slow. The quasi-equilibrium approximation itself describes slow motions. 
However, sometimes it becomes necessary to restore to the whole system, and to
take into account the fast motions as well. With this, it is desirable to keep intact 
one of the important advantages of the quasi-equilibrium approximation -
 its independence of the rate constants of fast reactions. 
For this purpose, the detailed fast kinetics is replaced by
a model equation ({\it single relaxation time approximation}).

{\it Quasi-equilibrium models} (QEM) are constructed as follows: For each concentration vector
$\cc$, consider the affine manifold, $\cc+L$. Its intersection with the quasi-equilibrium
manifold $\bOmega_L$ consists of one point. This point delivers the  minimum 
to $G$ on $\cc+L$. Let us denote this point as $\cc^*_L(\cc)$. The equation of the
quasi-equilibrium model reads:
\begin{equation}
\label{QEmodel}
\dot{\cc}=-\frac{1}{\tau}[\cc-\cc^*_L(\cc)]+\sum_{{\rm slow}}\bgamma_{s}W_s(\cc^*_L(\cc)),
\end{equation}
where $\tau>0$ is the relaxation time of the fast subsystem. Rates of slow reactions are 
computed in the points $\cc^*_L(\cc)$ (the second term in the 
right hand side of Eq.\ (\ref{QEmodel}), whereas the rapid motion is taken into account
by a simple relaxational term (the first term in the right hand side of Eq.\ (\ref{QEmodel}).
The most famous model kinetic equation is the BGK equation in the theory of the
Boltzmann equation \citep{BGK}. The general theory
of the quasi-equilibrium models, including proofs of their thermodynamic consistency,
was constructed by \cite{GK92c,GK94a}.

{\it Single relaxation time gradient models} (SRTGM) 
were considered by \cite{AK00,AK02,AK02a}
in the context of the lattice Boltzmann method for hydrodynamics. These models
are aimed at improving the obvious drawback of quasi-equilibrium models (\ref{QEmodel}):
In order to construct the QEM, one needs to compute 
the function,
\begin{equation}
\label{QEA}
 \cc^*_L(\cc)=\arg\min_{\xxx\in \cc+L,\ \xxx>0}G(\xxx).
\end{equation}
This is a convex programming problem. It does not always has a closed-form solution.

Let $\bg_1,\dots,\bg_k$ is the orthonormal basis of $L$. We denote 
as $\DD(\cc)$ the  $k\times k$ matrix with the elements 
$\langle \bg_i,\HH_{\cc}\bg_j\rangle$, where $\HH_{\cc}$ is the matrix of second derivatives
of $G$ (\ref{MATRIX}). Let $\CC(\cc)$ be the inverse of $\DD(\cc)$. The
single relaxation time gradient model has the form:
\begin{equation}
\dot{\cc}=-\frac{1}{\tau}\sum_{i,j}\bg_i\CC(\cc)_{ij}\langle \bg_j,\bnabla G\rangle
+\sum_{{\rm slow}}\bgamma_{s}W_s(\cc).\label{SRTGM}
\end{equation}
The first term drives the system to the minimum of $G$ on $\cc+L$, 
it does not require solving the problem (\ref{QEA}), and its 
spectrum in the quasi-equilibrium is the same as in the quasi-equilibrium model (\ref{QEmodel}).
Note that the slow component is evaluated in the ``current'' state $\cc$.

The models (\ref{QEmodel}) and (\ref{SRTGM}) 
 lift the quasi-equilibrium approximation to a kinetic equation
by approximating the fast dynamics with a single ``reaction rate constant'' -
relaxation time $\tau$.

\subsection{Quasi-steady state approximation}\label{QSS}
The quasi-steady state approximation (QSS) is a tool used in a huge amount of works.
Let us split the list of species in two groups: The basic and the intermediate (radicals etc).
Concentration vectors are denoted accordingly, $\cc^{\rm s}$ (slow, basic species),
and $\cc^{\rm f}$ (fast, intermediate species). The concentration vector
$\cc$ is the direct sum, $\cc=\cc^{\rm s}\oplus\cc^{\rm f}$.
The fast subsystem is Eq.\ (\ref{reaction}) for the component $\cc^{\rm f}$ at fixed
values of $\cc^{{\rm s}}$. If it happens that this way defined fast subsystem
relaxes to a stationary state, $\cc^{\rm f}\to\cc^{\rm f}_{\rm qss}(\cc^{\rm s})$, then the
assumption that $\cc^{\rm f}=\cc^{\rm f}_{\rm qss}(\cc)$ is precisely  the QSS assumption.
The slow subsystem is the part of the system (\ref{reaction}) for $\cc^{\rm s}$, in the
right hand side of which the component $\cc^{\rm f}$ is replaced with 
$\cc^{\rm f}_{\rm qss}(\cc)$. Thus, $\JJ=\JJ_{\rm s}\oplus\JJ_{\rm f}$, where
\begin{eqnarray}
\dot{\cc}^{\rm f}&=&\JJ_{\rm f}(\cc^{\rm s}\oplus\cc^{\rm f}), \ \cc^{\rm s}={\rm const};
\quad \cc^{\rm f}\to\cc^{\rm f}_{\rm qss}(\cc^{\rm s});\label{fast}\\
\dot{\cc}^{\rm s}&=&\JJ_{\rm s}(\cc^{\rm s}\oplus\cc_{\rm qss}^{\rm f}(\cc^{\rm s})).
\label{slow}
\end{eqnarray}
Bifurcations in the system (\ref{fast}) under variation of $\cc^{\rm s}$ as a parameter are 
confronted to kinetic critical phenomena. Studies of more complicated dynamic phenomena in 
the fast subsystem (\ref{fast}) require various techniques of averaging, stability analysis of the 
averaged quantities etc. 

Various versions of the QSS method are well possible, and are actually used widely,
for example, the hierarchical QSS method.
There, one defines not a single fast subsystem but a hierarchy of them,
$\cc^{{\rm f}_1},\dots,\cc^{{\rm f}_k}$. Each subsystem  $\cc^{{\rm f}_i}$ is regarded as 
a slow system for all the foregoing subsystems, and it is regarded as a fast subsystem for the
following members of the hierarchy. Instead of one system of equations (\ref{fast}),
a hierarchy of systems of lower-dimensional equations is considered, each of these
subsystem is easier to study analytically.

Theory of singularly perturbed systems of ordinary differential equations
is used to provide a mathematical background and further development of the
QSS approximation \citep{Bowen63,Segel89}.
In spite of a broad literature on this subject, it remains, in general, unclear, what 
is the smallness parameter that separates the intermediate (fast) species from
the basic (slow). Reaction rate constants cannot be such a parameter (unlike in the 
case of the quasi-equilibrium). Indeed, intermediate species participate 
in the {\it same} reactions, as the basic species (for example,
${\rm H}_2\rightleftharpoons 2{\rm H}$, ${\rm H}+{\rm O}_2\rightleftharpoons {\rm OH}+{\rm O}$).
It is therefore incorrect to state that $\cc^{\rm f}$ evolve faster than $\cc^{\rm s}$.
In the sense of reaction rate constants, $\cc^{\rm f}$ is not faster.

For catalytic reactions, it is not difficult to figure out what is the smallness parameter 
that separates the intermediate species from the basic, and which allows to
upgrade the QSS assumption to a singular perturbation theory rigorously
\citep{YBGE91}. This smallness parameter is the ratio of balances:
Intermediate species include the catalyst, and their total amount is simply
significantly less than the amount of all the $\cc_i$'s. After renormalizing to
the variables of one order of magnitude, the small parameter appears explicitly.

For usual radicals, the origin of the smallness parameter is quite similar.
There are much less radicals than the basic species (otherwise, the QSS
assumption is inapplicable). In the case of radicals, however, the smallness
parameter cannot be extracted directly from balances $B_i$ (\ref{conser}).
Instead, one can come up with a thermodynamic estimate: Function $G$
decreases in the course of reactions, whereupon we obtain the limiting estimate
of concentrations of any specie:
\begin{equation}
\label{TDlim}
c_i\le \max_{G(\cc)\le G(\cc(0))} c_i,
\end{equation}
where $\cc(0)$ is the initial composition. If the concentration $c_{\rm R}$ of the
radical R is small both initially and in the equilibrium, then it should remain small
also along the path to the equilibrium. For example, in the case of ideal $G$ (\ref{gfun})
under relevant conditions, for any $t>0$, the following inequality is valid:
\begin{equation}
\label{INEQ_R}
c_{\rm R}[\ln(c_{\rm R}(t)/c_{\rm R}^{\rm eq})-1]\le G(\cc(0)).
\end{equation}
Inequality (\ref{INEQ_R}) provides the simplest (but rather coarse) thermodynamic
estimate of $c_{\rm R}(t)$ in terms of $G(\cc(0))$ and $c_{\rm R}^{\rm eq}$
{\it uniformly for $t>0$}. Complete theory of thermodynamic estimates 
of dynamics has been developed by \cite{G84}.
One can also do computations without a priori estimations, if one accepts
the QSS assumption until the values $\cc^{\rm f}$ stay sufficiently small.

Let us assume that an a priori estimate has been found, $c_i(t)\le c_{i\ {\rm max}}$,
for each $c_i$. These estimate may depend on the initial conditions, thermodynamic data etc.
With these estimates, we are able to renormalize the variables in the kinetic equations
(\ref{reaction}) in such a way that renormalized variables take their values
from the unit segment $[0,1]$: $\tilde{c}_i=c_i/c_{i\ {\rm max}}$. Then the system (\ref{reaction})
can be written as follows:
\begin{equation}
\label{reduced}
\frac{d\tilde{c}_i}{dt}=\frac{1}{c_{i\ {\rm max}}}J_i(\cc).
\end{equation}
The system of dimensionless parameters, $\epsilon_i=c_{i\ {\rm max}}/\max_i c_{i\ {\rm max}}$
defines a hierarchy of relaxation times, and with its help one can establish various 
realizations of the QSS approximation. The simplest version is the standard QSS assumption:
Parameters $\epsilon_i$ are separated in two groups, the smaller ones, and of the order $1$.
Accordingly, the concentration vector is split into $\cc^{\rm s}\oplus\cc^{\rm f}$.
Various hierarchical QSS are possible, with this, the problem becomes more tractable
analytically.

Corrections to the QSS approximation can be addressed in various ways
(see, e.\ g., \cite{Vasil'eva95,Strygin88}). 
There exist a variety of ways to introduce the smallness parameter into kinetic equations,
and one can find applications to each of the realizations. However, the two particular
realizations remain basic for chemical kinetics:
(i) Fast reactions (under a given thermodynamic data);
(ii) Small concentrations. In the first case, one is led to the quasi-equilibrium approximation, 
in the second case - to the classical QSS assumption. Both of these approximations allow 
for hierarchical realizations,
those which include not just two but many relaxation time scales. Such a multi-scale approach
{\it essentially simplifies} analytical studies of the problem.

The method of invariant manifold which we present
below in the section \ref{MIM} allows to use both the QE and the QSS as initial approximations
in the iterational process of seeking slow invariant manifolds.
It is also possible
to use a  different initial ansatz chosen by a physical intuition, like, for example, the 
Tamm--Mott-Smith approximation in the theory of strong shock waves 
\citep{GK92}.

\subsection{Methods based on spectral decomposition of Jacobian fields}\label{Jacobi}

The idea to use the spectral decomposition of Jacobian fields in the problem 
of separating the motions into fast and slow originates from methods of analysis of
stiff systems \citep{Gear71}, and from methods of sensitivity analysis in control theory
\citep{Rabitz83}. There are two basic statements of the problem for 
these methods:
(i) The problem of the slow manifold, and
(ii) The problem of a complete decomposition (complete integrability) of kinetic equations.
The first of these problems consists in constructing the slow manifold $\bOmega$, and
a decomposition of motions into the fast one - towards $\bOmega$, and the slow
one - along $\bOmega$ \citep{Maas92}. The second of these problems consists
in a transformation of kinetic equations (\ref{reaction}) to a diagonal form,
$\dot{\zeta}_i=f_i(\zeta_i)$  (so-called
{\it full nonlinear lumping}
 or {\it modes decoupling}, \cite{Lam94,Li94,Toth97}).
 Clearly, if one finds a sufficiently explicit solution
to the second problem, then  the system (\ref{reaction}) is completely integrable, and nothing
more is needed, the result has to be simply used. The question is only to what extend
such a solution can be possible, and how difficult it would be as compared to the first
problem to find it. 

One of the currently  most popular methods is the construction of the so-called
{\it intrinsic low-dimensional manifold} (ILDM, \cite{Maas92}). 
This method is based on the following geometric picture: For each point $\cc$, 
one defines the Jacobian matrix of
Eq.\ (\ref{reaction}), $\FF_{\cc}\equiv \partial \JJ(\cc)/\partial\cc$. One assumes that, in the 
domain of interest, the eigenvalues of $\FF_{\cc}$ are separated into two groups, 
$\lambda_i^{\rm s}$ and $\lambda_j^{\rm f}$, and that the following inequalities are valid:
\[ {\rm Re}\ \lambda_i^{\rm s}\ge a > b\ge {\rm Re} \lambda_j^{\rm f},\ a\gg b,\ b<0.\]
Let us denote as $L_{\cc}^{\rm s}$ and $L_{\cc}^{\rm f}$ the invariant subspaces corresponding
to $\lambda^{\rm s}$ and $\lambda^{\rm f}$, respectively, and let $\ZZ_{\cc}^{\rm s}$ 
and $\ZZ_{\cc}^{\rm f}$ be the corresponding spectral projectors,
$\ZZ_{\cc}^{\rm s}L_{\cc}^{\rm s}=L_{\cc}^{\rm s}$, 
$\ZZ_{\cc}^{\rm f}L_{\cc}^{\rm f}=L_{\cc}^{\rm f}$,
$\ZZ_{\cc}^{\rm s}L_{\cc}^{\rm f}=\ZZ_{\cc}^{\rm f}L_{\cc}^{\rm s}=\{0\}$,
$\ZZ_{\cc}^{\rm s}+\ZZ_{\cc}^{\rm f}=1$.
Operator $\ZZ_{\cc}^{\rm s}$ projects onto the subspace of ``slow modes'' $L_{\cc}^{\rm s}$,
and it annihilates the ``fast modes'' $L_{\cc}^{\rm f}$. Operator
$\ZZ_{\cc}^{\rm f}$ does the opposite, it projects onto fast modes, and it annihilates
the slow modes. The basic equation of the ILDM reads:
\begin{equation}
\label{ILDM}
\ZZ_{\cc}^{\rm f}\JJ(\cc)=0.
\end{equation}
In this equation, the unknown is the concentration vector $\cc$. The set of solutions 
to Eq.\ (\ref{ILDM}) is the ILDM manifold $\bOmega_{{\rm ildm}}$.

For linear systems, $\FF_{\cc}$, $\ZZ_{\cc}^{\rm s}$, and $\ZZ_{\cc}^{\rm f}$, do not depend
on $\cc$, and $\bOmega_{{\rm ildm}}=\cc^{\rm eq}+L^{\rm s}$. On the other hand, obviously,
$\cc^{\rm eq}\in\bOmega_{{\rm ildm}}$. Therefore, procedures of solving of Eq.\ (\ref{ILDM})
can be initiated by choosing the linear approximation, 
$\bOmega_{{\rm ildm}}^{(0)}=\cc^{\rm eq}+L^{\rm s}_{\cc^{\rm eq}}$, 
in the neighborhood of the equilibrium 
$\cc^{\rm eq}$, and then continued parametrically into the nonlinear domain. 
Computational technologies of a continuation of solutions with respect to
parameters are well developed (see, for example, \cite{Khibnik93,Roose90}).
The problem of the relevant parameterization is solved locally: In the neighborhood of
a given point $\cc^0$ one can choose $\ZZ_{\cc}^{\rm s}(\cc-\cc^0)$ for a characterization
of the vector $\cc$. In this case, the space of parameters is $L_{\cc}^{\rm s}$.
There exist other, physically motivated ways to parameterize manifolds 
(\cite{GK92}; see also section \ref{TDP} below). 

There are two drawbacks of the ILDM method which call for its refinement:
(i) {\it ``Intrinsic'' does not imply ``invariant''.} Eq.\ (\ref{ILDM})
is not invariant of the dynamics (\ref{reaction}). If one differentiates Eq.\ (\ref{ILDM})
in time due to Eq.\ (\ref{reaction}), one obtains a new equation which is the
implication of Eq.\ (\ref{ILDM}) {\it only} for linear systems. In a general case, the
motion $\cc(t)$ takes off the $\bOmega_{{\rm ildm}}$. 
Invariance of a manifold $\bOmega$ means that $\JJ(\cc)$ touches $\bOmega$ in every
point $\cc\in\bOmega$. It remains unclear how the ILDM (\ref{ILDM}) corresponds with
this condition. Thus, from the dynamical perspective, the status of the
ILDM remains not well defined, or ``ILDM is ILDM'', defined self-consistently
by Eq.\ (\ref{ILDM}), and that is all what can be said about it.
(ii) From the geometrical standpoint, spectral decomposition of Jacobian fields
is not the most attractive way to compute manifolds. 
If we are interested in the behavior of trajectories, how they converge or diverge,
then one should consider the symmetrized part of $\FF_{\cc}$, rather than
$\FF_{\cc}$ itself.

Symmetric part, $\FF_{\cc}^{\rm sym}=(1/2)(\FF_{\cc}^{\dag}+\FF_{\cc})$,
defines the dynamics of the distance between two solutions,  $\cc$ and $\cc'$,
 in a given local Euclidean metrics.
Skew-symmetric part defines rotations.
If we want to study manifolds based on the argument about convergence/divergence
of trajectories, then we should use in Eq.\ (\ref{ILDM}) 
the spectral projector $\ZZ_{\cc}^{\rm f sym}$ for the
operator $\FF_{\cc}^{\rm sym}$. This, by the way, is  also a  significant simplification
from the standpoint of computations.
It remains to choose the metrics.
This choice is unambiguous from the thermodynamic perspective.
In fact, there is only one choice which fits into the physical meaning of the problem,
this is the metrics associated with the  thermodynamic (or entropic) scalar product,
\begin{equation}
\label{ESP}
\langle\langle\xxx,\yy\rangle\rangle=\langle\xxx,\HH_{\cc}\yy\rangle,
\end{equation}
where $\HH_{\cc}$ is the matrix of second-order derivatives
of $G$ (\ref{MATRIX}). In the equilibrium, operator $\FF_{\cc^{\rm eq}}$ is
selfajoint with respect to this scalar product (Onsager's reciprocity relations).
Therefore, the behavior of the ILDM in the vicinity of the equilibrium does 
not alter under the replacement, $\FF_{\cc^{\rm eq}}=\FF_{\cc^{\rm eq}}^{\rm sym}$.
In terms of usual matrix representation, we have:
\begin{equation}
\FF_{\cc}^{\rm sym}=\frac{1}{2}(\FF_{\cc}+\HH_{\cc}^{-1}\FF_{\cc}^{T}\HH_{\cc}),\label{Fsym}
\end{equation}
where $\FF_{\cc}^{T}$ is the ordinary transposition. 

The ILDM constructed with the help of the symmetrized Jacobian field will be termed
the {\it symmetric entropic intrinsic low-dimensional manifold} (SEILDM).
Selfadjointness of $\FF_{\cc}^{\rm sym}$ (\ref{Fsym}) with respect to the
thermodynamic scalar product (\ref{ESP}) simplifies considerably computations of 
 spectral decomposition. Moreover, it becomes necessary to do spectral decomposition
in only one point - in the equilibrium. Perturbation theory for selfadjoint operators
is a very well developed subject \citep{Kato76}, which makes it possible to easily extend the
spectral decomposition with respect to parameters.
A more detailed discussion of the selfadjoint linearization will be given below in section
\ref{SA}. 

Thus, when the geometric picture behind the decomposition of motions is specified,
the physical significance of the ILDM becomes more transparent, and it leads to
its modification into the SEILDM. This also gains simplicity in the implementation
by switching from non-selfadjoint spectral problems to selfadjoint. The
quantitative estimate of this simplification is readily available: Let $d$ be the dimension
of the phase space, and $k$ the dimension of the ILDM ($k={\rm dim} L_{\cc}^{\rm s}$).
The space of all the projectors $\ZZ$ with the $k$-dimensional image
has the dimension $D=2k(d-k)$. The space of all the selfadjoint projectors with the
$k$-dimensional image has the dimension $D^{\rm sym}=k(d-k)$.
For $d=20$ and $k=3$, we have $D=102$ and $D^{\rm sym}=51$.
When the spectral decomposition by means of parametric extension is addressed,
one considers equations of the form:
\begin{equation}
\label{parametric}
\frac{d\ZZ_{\cc(\tau)}^{\rm s}}{d\tau}=\bPsi^{\rm s}\left(\frac{d\cc}{d\tau},
\ZZ_{\cc(\tau)}^{\rm s}, \FF_{\cc(\tau)}, \bnabla\FF_{\cc(\tau)}\right),
\end{equation}
where $\tau$ is the parameter, and $\bnabla\FF_{\cc}=\bnabla\bnabla\JJ(\cc)$
is the differential of the Jacobian field. 
For the selfadjoint case, where we use $=\FF_{\cc}^{\rm sym}$ instead of
 $\FF_{\cc}$, this system of equations
has twice less independent variables, and also the right hand is of a simpler structure.

It is more difficult to improve on  the first of the remarks (ILDM is not invariant).
The following naive approach may seem possible: 

(i) Take $\bOmega_{\rm ildm}=\cc^{\rm eq}+L^{\rm s}_{\cc^{\rm eq}}$ in a neighborhood
$U$ of the equilibrium $\cc^{\rm eq}$. [This is also a useful initial approximation
for solving Eq.\ (\ref{ILDM})].

(ii) Instead of computing the solution to Eq.\ (\ref{ILDM}), integrate the kinetic
equations (\ref{reaction}) {\it backwards in the time}. It is sufficient to take
initial conditions $\cc(0)$ from a dense set on the boundary,
$\partial U\cap (\cc^{\rm eq}+L^{\rm s}_{\cc^{\rm eq}})$, and to compute
solutions $\cc(t)$, $t<0$.

(iii) Consider the obtained set of trajectories as an approximation of the slow invariant manifold.

This approach will guarantee invariance, by construction, but it is prone to pitfalls in what 
concerns the slowness. Indeed, the integration backwards in the time will see
exponentially divergent trajectories, if they were exponentially converging in the
normal time progress. This way one finds {\it some} invariant manifold which touches
$\cc^{\rm eq}+L^{\rm s}_{\cc^{\rm eq}}$ in the equilibrium. Unfortunately, there are infinitely
many such manifolds, and they fill out almost all the space of concentrations.
However, we must select the slow component of motions. Such a regularization is possible.
Indeed, let us replace in Eq.\ (\ref{reaction}) the vector field $\JJ(\cc)$
by the vector field $\ZZ_{\cc}^{{\rm s sym}}\JJ(\cc)$, and obtain a regularized
kinetic equation,
\begin{equation}
\label{regreaction}
\dot{c}=\ZZ_{\cc}^{{\rm s sym}}\JJ(\cc).
\end{equation}
Let us replace integration backwards in time of the kinetic equation (\ref{reaction})
in the naive approach described above by integration backwards in time of the
regularized kinetic equation (\ref{regreaction}). With this, we obtain a rather
convincing version of the ILDM (SEILDM).
Using Eq.\ (\ref{parametric}), one also can write down an
equation for the projector $\ZZ_{\cc}^{{\rm s sym}}$, putting $\tau=t$.
Replacement of Eq.\ (\ref{reaction}) by Eq.\ (\ref{regreaction})
also makes the integration backwards in time in the naive approach more stable.
However, {\it regularization will again conflict with invariance}.
The ``naive refinement'' after the regularization (\ref{regreaction})
produces just a slightly  different  version of the ILDM (or SEILDM) but
it does not construct the slow invariant manifold.
So, where is the way out? We believe that the ILDM and its version SEILDM
are, in general, good initial approximations of the slow manifold. However,
if one is indeed interested in finding the invariant manifold, one has to write 
out the true condition of invariance and solve it. As for the initial approximation
for the method of invariant manifold one can use any ansatz, in particular,
the SEILDM.

{\it The problem of a complete decomposition} of kinetic equations can be solved indeed in
some cases. The first such solution was the spectral decomposition for linear systems
\citep{Wei62}. Decomposition is sometimes possible also for 
nonlinear systems (\cite{Li94}; \cite{Toth97}). 
The most famous example of a complete decomposition 
of infinite-dimensional kinetic equation is the complete integrability
of the space-independent Boltzmann equation for Maxwell`s molecules found
by \cite{Bobylev88}.
However, in a general case, there exist no analytical, not even a twice differentiable
transformation which would decouple modes. The well known Grobman-Hartman 
theorem \citep{Hartman63,Hartman82} 
states only the existence of a continuous transform which decomposes 
modes in a neighborhood of the equilibrium.
For example, the analytic planar system,
$dx/dt=-x$, $dy/dt=-2y+x^2$, is not $C^2$ linearizable. These problems remain of interest
 \citep{Chicone00}.
Therefore, in particular, it becomes quite ineffective to construct such a transformation in a 
form of a series.
It is more effective to solve a simpler problem of extraction of a slow invariant manifold
\citep{Beyn98}. 

{\it Sensitivity analysis} \citep{Rabitz83,Rabitz87,Lam94}
makes it possible to select essential variables and reactions, and to decompose motions into
fast  and slow. In a sense, the ILDM method is a development
of the sensitivity analysis. Recently, a further step in this direction was done by 
\cite{Zhu99}. In this work, the authors use a {\it nonlocal in time criterion of 
closeness of solutions} of the full and of the reduced systems of chemical kinetics.
They require not just a closeness of derivatives but a true closeness of the dynamics.

Let us be interested in the dynamics of the concentrations of just a few species,
${\rm A}_1,\dots, {\rm A}_{p}$, whereas the rest of the species, ${\rm A}_{p+1},
\dots, {\rm A}_{n}$ are used for building the kinetic equation, and for understanding
the process. Let $\cc_{\rm goal}$ be the concentration vector with components
$c_1,\dots, c_p$, $\cc_{\rm goal}(t)$ be the corresponding components of the 
solution to Eq.\ (\ref{reaction}),
and $\cc^{\rm red}_{\rm goal}$ be the solution to the simplified model with
corresponding initial conditions. 
\cite{Zhu99} suggest to minimize the difference between 
 $\cc_{\rm goal}(t)$ and  $\cc^{\rm red}_{\rm goal}$ on the
segment $t\in[0,T]$: $\|\cc_{\rm goal}(t)-\cc^{\rm red}_{\rm goal}\|\to\min$.
In the course of the optimization under certain restrictions one
selects the optimal (or appropriate) reduced model.
The sequential quadratic programming method and heuristic rules of 
sorting the reactions, substances etc were used.
In the result, for some stiff systems studied, one avoids typical pitfalls
of the local sensitivity analysis. In simpler situations this method should give 
similar results as the local methods.

\subsection{Thermodynamic criteria for selection of important reactions}
One of the problems addressed by the sensitivity analysis is the selection of the important
and discarding the unimportant reactions. \cite{BYA77}
suggested a simple principle to compare importance of different reactions according
to their contribution to the entropy production (or, which is the same, according to
their contribution to $\dot{G}$). Based on this principle, \cite{Dimitrov82}  described domains
of parameters in which the reaction of hydrogen oxidation, ${\rm H}_2+{\rm O}_2+{\rm M}$,
proceeds due to different mechanisms. For each elementary reaction, he has derived the domain
inside which the contribution of this reaction is essential (nonnegligible).
Due to its simplicity, this entropy production principle is especially well suited
for analysis of complex problems.
In particular, recently, a version of the entropy production principle
was used in the problem of selection of boundary conditions for Grad's moment equations
\citep{Struchtrup98,GKZ02}. For ideal systems (\ref{gfun}), the contribution of the
$s$th reaction to $\dot{G}$ has a particularly simple form:
\begin{equation}
\label{dotGs}
\dot{G}_{s}=-W_s\ln\left(\frac{W_s^+}{W_s^-}\right),\ \dot{G}=\sum_{s=1}^{r}\dot{G}_s.
\end{equation}
For nonideal systems, the corresponding expressions (\ref{HMDD})
 are also not too complicated.

\section{Outline of the method of invariant manifold}
\label{MIMG}
In many cases, dynamics of the $d$-dimensional system (\ref{reaction})
leads to a manifold of a lower dimension. Intuitively, a typical
phase trajectory behaves as follows: Given the initial state
$\cc(0)$ at $t=0$, and after some period of time, the trajectory comes
close to some low-dimensional manifold $\bOmega$, and after that 
proceeds towards the equilibrium essentially along this manifold.
The goal is to construct this manifold.

The starting point of our approach is based on a formulation of
the two main requirements:

(i). {\it Dynamic invariance}:  The manifold $\bOmega$
should  be (positively)  invariant  under the dynamics  of the originating system
(\ref{reaction}): If $\cc(0)\in\bOmega$, then $\cc(t)\in\bOmega$ for
each $t>0$.

(ii). {\it Thermodynamic consistency of the reduced dynamics}:
Let  {\it some} (not obligatory invariant)
manifold $\bOmega$ is considered as a manifold of
reduced description. We should define a set of linear operators,
$\PP_{\cc}$, labeled by the states $\cc\in\bOmega$, which
project the vectors $\JJ(\cc)$, $\cc\in\bOmega$ onto the
tangent bundle of the manifold $\bOmega$, thereby
generating the induced vector field,
$\PP_{\cc}\JJ(\cc)$, $\cc\in\bOmega$.
This induced vector field on the tangent bundle of the manifold
$\bOmega$ is identified with the reduced dynamics
along the manifold $\bOmega$.
The thermodynamicity requirement for this induced vector field reads
\begin{equation}
\label{thermo}
\langle\bnabla G(\cc),\PP_{\cc}\JJ(\cc)\rangle\leq 0,\mbox{ for\ each\ }\cc\in\bOmega.
\end{equation}

In order to meet these requirements, the method of invariant
manifold suggests two complementary procedures:

(i). To treat the condition of dynamic invariance as an equation,
and to solve it iteratively by a Newton method.
This procedure is geometric in its nature, and it does not
use the time dependence and small parameters.

(ii). Given an approximate manifold of reduced description,
to construct the projector satisfying the condition
(\ref{thermo}) in a way which does not depend on the vector
field $\JJ$.

We shall now outline both these procedures starting with the second.
The solution consists, in the first place, in formulating the
{\it thermodynamic condition} which should
be met by the projectors $\PP_{\cc}$:
For each $\cc\in\bOmega$, let us consider the linear functional
\begin{equation}
\label{FUNC}
M^*_{\cc}(\xx)=\langle\bnabla G(\cc),\xx\rangle.
\end{equation}
Then the thermodynamic condition for the projectors reads:
\begin{equation}
\label{therm1}
{\rm ker}\PP_{\cc}\subseteq{\rm ker}M^*_{\cc},\ 
{\rm  for\ each}\ \cc\in\bOmega.
\end{equation}
Here ${\rm ker}\PP_{\cc}$ is the null space of the projector, and
${\rm ker}M^*_{\cc}$ is the hyperplane orthogonal to the vector
$M^*_{\cc}$.
It has been shown  \citep{GK92,GK94}
that the condition (\ref{therm1}) is
the necessary and sufficient condition to establish the thermodynamic
induce vector field on the given manifold $\bOmega$ for all possible
dissipative vector fields $\JJ$ simultaneously.

Let us now turn to the requirement of invariance.
By a definition, the manifold  $\bOmega$  is  invariant
with respect to the vector field $\JJ$  if and only if 
the following equality is true:
\begin{equation}
\label{INVARIANCE}
\left[1-\PP\right]\JJ(\cc)=0,\ {\rm for\ each\ } \cc\in\bOmega.
\end{equation}
In this expression $\PP$ is an {\it arbitrary}
projector on the tangent
bundle of the manifold $\bOmega$. 
It has been suggested to consider the condition (\ref{INVARIANCE}) as an
{\it equation} to be solved iteratively starting with some appropriate initial
manifold.

Iterations for the invariance equation (\ref{INVARIANCE})
are considered  in the section \ref{DC}.
The next section presents construction of the thermodynamic projector
using a specific parameterization of manifolds.

\section{Thermodynamic projector}
\label{THERMO}
\subsection{Thermodynamic parameterization}
\label{TDP}
In this section, $\bOmega$ denotes a generic $p$--dimensional
manifold. First, it should be mentioned that {\it any}
parameterization of $\bOmega$ generates a certain projector, and
thereby a certain reduced dynamics.
Indeed, let us consider a set of $m$ independent functionals
$M(\cc)=\{M_1(\cc),\dots,M_p(\cc)\}$,
and let us assume that they form a
coordinate system on $\bOmega$ in such a way that
$\bOmega=\cc(M)$, where $\cc(M)$ is a vector function of the parameters
$M_1,\dots,M_p$. Then the projector associated with
this parameterization reads:
\begin{equation}
\label{proj}
\PP_{\cc(M)}\xx=\sum_{i,j=1}^p\frac{\partial\cc(M)}{\partial M_i}
N^{-1}_{ij}(M)
\langle\bnabla M_j\bigm|_{\cc(M)},\xx\rangle,
\end{equation}
where $N^{-1}_{ij}$ is the inverse to the $p\times p$
matrix:
\begin{equation}
\label{NORMALIZATION}
\NN(M)=\|\langle \bnabla M_i,\partial\cc/\partial M_j\rangle\|.
\end{equation}
This somewhat involved notation is intended to stress that the projector
(\ref{proj}) is dictated by the choice of the parameterization.
Subsequently, the induced vector field of the reduced dynamics
is found by applying projectors (\ref{proj}) on the vectors
$\JJ(\cc(M))$, thereby inducing the reduced dynamics in terms
of the parameters $M$ as follows:
\begin{equation}
\label{dyn}
\dot{M}_i=\sum_{j=1}^pN^{-1}_{ij}(M)
\langle\bnabla M_j\bigm|_{\cc(M)},\JJ(\cc(M))\rangle,
\end{equation}
Depending on the choice of the parameterization,
dynamic equations (\ref{dyn}) are (or are not) consistent
with the thermodynamic requirement (\ref{thermo}).
The {\it thermodynamic parameterization} makes use of the condition
(\ref{therm1}) in order to establish the thermodynamic projector.
Specializing to the case (\ref{proj}), let us consider the linear functionals,
\begin{equation}
\label{derivative}
DM_i\bigm|_{\cc(M)}(\xx)=
\langle\bnabla M_i\bigm|_{\cc(M)},\xx\rangle.
\end{equation}
Then the condition (\ref{therm1}) takes the form:
\begin{equation}
\label{therm2}
\bigcap_{i=1}^p{\rm ker}DM_i\bigm|_{\cc(M)}\subseteq{\rm ker}
M^*_{\cc(M)},
\end{equation}
that is, the intersection of null spaces of the functionals (\ref{derivative})
should belong to the null space of the differential of the
Lyapunov function $G$, in each point of the manifold $\bOmega$.

In practice, in order to construct the thermodynamic
parameterization, we take the following set of
functionals in each point $\cc$ of the manifold $\bOmega$: 
\begin{eqnarray}
\label{param_thermo}
M_1(\xx)&=&M_{\cc}^*(\xx),\ \cc\in\bOmega\\
M_i(\xx)&=&\langle\mm_i,\xx\rangle,\ i=2,\dots,p
\label{param_other}
\end{eqnarray}
It is required that vectors $\bnabla G(\cc), \mm_2,\dots,\mm_p$
are linearly independent in each state $\cc\in\bOmega$.
Inclusion of the functionals (\ref{FUNC}) as a part of the 
system (\ref{param_thermo}) and (\ref{param_other}) 
implies the thermodynamic condition
(\ref{therm2}). Also, any linear combination of the parameter
set (\ref{param_thermo}), (\ref{param_other})  will meet the thermodynamicity requirement.

It is important to notice here that the thermodynamic
condition is satisfied whatsoever the functionals $M_2,\dots,M_p$ are.
This is very convenient for it gives an opportunity to take into account
the conserved quantities correctly.
The manifolds we are going to deal with should be consistent
with the conservation laws (\ref{conser}). While the explicit
characterization of the phase space $\VV$ is a problem on its own,
in practice, it is customary to work in the $n$--dimensional space 
while keeping the constraints (\ref{conser}) explicitly on each step
of the construction. For this technical reason, it is convenient to 
consider manifolds of the dimension $p>l$, where $l$ is the number of conservation 
laws, in the $n$--dimensional space rather than in the phase space
$\VV$.  The thermodynamic parameterization is then concordant also
with the conservation laws if $l$ of the linear functionals
(\ref{param_other}) are identified with the conservation laws.
In the sequel, only projectors consistent
with conservation laws are considered.

Very frequently, the manifold $\bOmega$ is represented as a 
$p$-parametric family $\cc(a_1,\dots,a_p)$, where
$a_i$ are coordinates on the manifold. The thermodynamic 
{\it re-parameterization} 
suggests a representation of the coordinates $a_i$ in terms of
$M_{\cc}^*, M_2,\dots,M_p$ (\ref{param_thermo}), (\ref{param_other}).
While the explicit construction of these functions may be
a formidable task, we notice that the construction
of the thermodynamic projector of the form (\ref{proj}) and of the
dynamic equations (\ref{dyn}) is relatively easy because only
the derivatives $\partial\cc/\partial M_i$ enter these expressions.
This point was discussed in a detail by  \cite{GK92,GK94}.

\subsection{Decomposition of motions: Thermodynamics}
\label{DEC_THERM}
Finally, let us discuss how the thermodynamic projector is related to the
decomposition of motions.
{\it Assuming} that the decomposition of motions near the manifold
$\bOmega$ is true indeed, let us consider states which were initially
close enough to the manifold $\bOmega$. Even without knowing
the details about the evolution of the states towards $\bOmega$,
we know that the Lyapunov function $G$ was decreasing in the
course of this evolution. Let us consider a set of states $\UU_{\cc}$
 which contains all those vectors $\cc'$ that 
have arrived (in other words, have been projected) into the point
$\cc\in\bOmega$. Then we observe that the
state $\cc$ furnishes the minimum of  the function $G$ on the set $\UU_{\cc}$.
 If a state $\cc'\in\UU_{\cc}$, and if it deviates small enough
from the state $\cc$ so that the linear
approximation is valid, then $\cc'$ belongs to the affine hyperplane
\begin{equation}
\label{hyperplane}
\Gamma_{\cc}=\cc+{\rm ker\ }M_{\cc}^*,\ \cc\in\bOmega.
\end{equation}
This hyperplane actually participates in the condition (\ref{therm1}).
The consideration was entitled `thermodynamic' \citep{GK92}
because it describes the states $\cc\in\bOmega$ as points of minimum
of the function $G$ over the corresponding hyperplanes
(\ref{hyperplane}).

\section{Corrections}
\label{DC}
\subsection{Preliminary discussion}
The thermodynamic projector is needed to induce the dynamics on a
given manifold in such a way that the dissipation inequality (\ref{thermo})
holds. Coming back to the issue of constructing corrections, we should
stress that the projector participating in the invariance condition
(\ref{INVARIANCE}) is arbitrary. It is convenient to make
use of this point: When Eq.\ (\ref{INVARIANCE}) is solved iteratively, the
projector may be kept non--thermodynamic unless the induced dynamics
is explicitly needed.

Let us assume that we have chosen
the initial manifold, $\bOmega_0$, together with the associated
projector $\PP_0$, as the first approximation to the
desired manifold of reduced description. Though the choice
of the initial approximation $\bOmega_0$ depends on the specific
problem, it is often reasonable to consider quasi-equilibrium or
quasi steady-state approximations.
In most cases, the manifold $\bOmega_0$ is not
an invariant manifold. This means that $\bOmega_0$ does not
satisfy the invariance condition
(\ref{INVARIANCE}):
\begin{equation}
\label{DEFECT}
\bDelta_0=[1-\PP_0]\JJ(\cc_0)\ne0,\
{\rm for\ some\ } \cc_0\in\bOmega_0.
\end{equation}
Therefore, we seek a correction $\cc_1=\cc_0+\delta\cc$.
Substituting $\PP=\PP_0$ and $\cc=\cc_0+\delta\cc$ into the
invariance equation (\ref{INVARIANCE}), and after the linearization in
$\delta\cc$, we derive the following linear equation:
\begin{equation}
\label{METHOD}
\left[1-\PP_{0}\right]\left[\JJ(\cc_0)+
\LL_{\cc_0}\delta\cc\right]=0,
\end{equation}
where $\LL_{c_0}$ is the matrix of first derivatives of the
vector function $\JJ$, computed in the state $\cc_0\in\bOmega_0$.
The system of linear algebraic equations (\ref{METHOD}) should be
supplied with the additional condition. 
\begin{equation}
\label{uni_con}
\PP_0\delta\cc=0.
\end{equation}

In order to illustrate the nature of the Eq.\ (\ref{METHOD}), let us consider
the case of linear manifolds for linear systems. 
Let a linear evolution equation is given in the finite-dimensional real space: 
$\dot{\cc}=\LL\cc$, where $\LL$ is negatively definite symmetric matrix
with a simple spectrum.
Let us further assume the quadratic Lyapunov function,
$G(\cc)=\langle\cc,\cc\rangle$. The manifolds we consider are lines,
$\vl(a)=a\ee$, where $\ee$ is the unit vector, and $a$ is a scalar.
The invariance equation for such manifolds reads:
$\ee\langle\ee,\LL\ee\rangle-\LL\ee=0$,
and is simply a form of the eigenvalue problem for the operator $\LL$.
Solutions to the latter equation are eigenvectors $\ee_{i}$, 
corresponding to eigenvalues $\lambda_{i}$.

Assume that we have chosen a line, $\vl_0=a\ee_0$, defined by the unit
vector $\ee_0$, and that $e_0$ is not an eigenvector of $\LL$.
We seek another line, $\vl_1=a\ee_1$, where $\ee_1$ is another unit
vector, $\ee_1=\yy_1/\|\yy_1\|$, $\yy_1=\ee_0+\delta\yy$. The additional
condition (\ref{uni_con}) now reads: $\langle\delta\yy,\ee_0\rangle=0$.
Then the Eq.\ (\ref{METHOD}) becomes
$[1-\ee_0\langle\ee_0,\cdot\rangle]L[\ee_0+\delta\yy]=0$.
Subject to the additional condition, the
unique solution is as follows: $\ee_0+\delta\yy=
\langle\ee_0,\LL^{-1}\ee_0\rangle^{-1}\LL^{-1}\ee_0$.
Rewriting the latter expression in the eigen--basis of $\LL$,
we have: $\ee_0+\delta\yy\propto
\sum_{i}\lambda_i^{-1}\ee_i\langle\ee_i,\ee_0\rangle$.
The leading term in this sum corresponds to the 
eigenvalue with the minimal absolute value.
The example indicates that 
the  method of linearization (\ref{METHOD}) seeks  
the direction of the {\it slowest relaxation}.
For this reason, the method (\ref{METHOD}) can be
recognized as the basis of an iterative method for 
constructing the manifolds of slow motions. 

For the nonlinear systems, the matrix $\LL_{c_0}$ in the Eq.\ (\ref{METHOD}) 
depends nontrivially on $\cc_0$. In this case
the system (\ref{METHOD}) requires a further specification
which will be done now.

\subsection{Symmetric linearization}
\label{SA}
The invariance condition
(\ref{INVARIANCE}) supports a lot of invariant manifolds, and not all of
them are relevant to the reduced description
(for example, any individual trajectory is itself an invariant manifold).
This should be carefully taken into
account when deriving a relevant equation for the correction in
the states of the initial manifold $\bOmega_0$ which are located far from
equilibrium. This point concerns the procedure of the linearization of
the vector field $\JJ$, appearing in the
equation (\ref{METHOD}).
We shall return to the explicit form of the Marcelin--De Donder kinetic
function (\ref{MDD}). Let $\cc$ is an arbitrary fixed element of
the phase space. The linearization of the vector function $\JJ$
(\ref{KINETIC MDD}) about $\cc$ may be written
$\JJ(\cc+\delta\cc)\approx\JJ(\cc)+\LL_{\cc}\delta\cc$
where the linear operator $\LL_{\cc}$ acts as follows:
\begin{equation}
\label{LINEAR}
\LL_{\cc}\xx=\sum_{s=1}^{r}
\bgamma_s[W^{+}_s(\cc)
\langle\balpha_s,\HH_{\cc}\xx
\rangle-
W^{-}_s(\cc)\langle\bbeta_s,\HH_{\cc}
\xx\rangle].
\end{equation}
Here $\HH_{\cc}$ is the matrix of second derivatives of the
function $G$ in the state $\cc$ [see Eq.\ (\ref{MATRIX})].
The matrix $\LL_{\cc}$ in the Eq.\ (\ref{LINEAR}) can be decomposed
as follows:
\begin{equation}
\label{DECOMPOSITION}
\LL_{\cc}=\LL'_{\cc}+
\LL''_{\cc}.
\end{equation}
Matrices $\LL'_{\cc}$ and $\LL''_{\cc}$ act as follows:
\begin{eqnarray}
\label{SYMMETRIC}
\LL'_{\cc}\xx&=&-\frac{1}{2}\sum_{s=1}^{r}
[W^{+}_s(\cc)+W^{-}_s(\cc)]
\bgamma_s\langle\bgamma_s,
\HH_{\cc}\xx\rangle,\\
\label{NONSYMMETRIC}
\LL''_{\cc}\xx&=&\frac{1}{2}\sum_{s=1}^{r}
[W^{+}_s(\cc)-W^{-}_s(\cc)]\bgamma_s
\langle\balpha_s+\bbeta_s,
\HH_{\cc}\xx\rangle.
\end{eqnarray}
Some features of this decomposition are best seen when we
use the thermodynamic scalar product (\ref{ESP}):
The following properties of the matrix $\LL'_{\cc}$ are verified
immediately:

(i) The matrix $\LL'_{\cc}$ is symmetric in the scalar product
(\ref{ESP}):
\begin{equation}
\label{sym}
\langle\!\langle\xx,\LL'_{\cc}\yy\rangle\!\rangle=
\langle\!\langle\yy,\LL'_{\cc}\xx\rangle\!\rangle.
\end{equation}
(ii) The matrix $\LL'_{\cc}$ is nonpositive definite in the scalar product
(\ref{ESP}):
\begin{equation}
\label{pos}
\langle\!\langle\xx,\LL'_{\cc}\xx\rangle\!\rangle\le 0.
\end{equation}
(iii) The null space of the matrix $\LL'_{\cc}$ 
is the linear envelope of the vectors $\HH^{-1}_{\cc}\bb_i$ 
representing the complete system of conservation laws:
\begin{equation}
\label{bal}
{\rm ker}\LL'_{\cc}={\rm Lin}\{\HH^{-1}_{\cc}\bb_i, i=1,\dots,l\}
\end{equation}
(iv)  If $\cc=\cc^{\rm eq}$, then 
$W_s^+(\cc^{\rm eq})=W_s^-(\cc^{\rm eq})$, 
and 
\begin{equation}
\label{prop4}
\LL'_{\cc^{\rm eq}}=\LL_{\cc^{\rm eq}}.
\end{equation}

Thus, the decomposition Eq.\ (\ref{DECOMPOSITION}) splits the
matrix $\LL_{\cc}$ in two parts: one part, Eq.\ (\ref{SYMMETRIC})
is symmetric and nonpositive
definite, while the other part, Eq.\ (\ref{NONSYMMETRIC}),
vanishes in the equilibrium. The
decomposition Eq.\ (\ref{DECOMPOSITION}) explicitly takes into account
the Marcelin-De Donder form of the kinetic function.
For other dissipative systems, the decomposition
(\ref{DECOMPOSITION}) is possible as soon as 
the relevant kinetic operator is written in a gain--loss form [for instance,
this is straightforward for the Boltzmann collision operator].

In the sequel, we shall make use of the properties of the operator
$\LL'_{\cc}$ (\ref{SYMMETRIC}) for constructing the dynamic correction
by extending the picture of the decomposition of motions.

\subsection{Decomposition of motions: Kinetics}
\label{DEC_KIN}
The assumption about the existence of the decomposition of motions
near the manifold of reduced description $\bOmega$ has led to the
{\it thermodynamic} specifications of the states $\cc\in\bOmega$.
This was accomplished in the section \ref{DEC_THERM}, where the
thermodynamic projector was backed by an appropriate
variational formulation, and this helped us to establish
the induced dynamics consistent with the dissipation property.
Another important feature of the decomposition of motions is that the
states $\cc\in\bOmega$ can be specified {\it kinetically}.
Indeed, let us do it again as if the decomposition of motions
were valid in the neighborhood of the manifold $\bOmega$, and let us
 `freeze' the slow dynamics along the $\bOmega$, focusing on the 
fast process of relaxation
towards a state $\cc\in\bOmega$. From the thermodynamic perspective,
fast motions take place on the affine hyperplane 
$\cc+\delta\cc\in\Gamma_{\cc_0}$, where $\Gamma_{\cc_0}$ is given by Eq.\ (\ref{hyperplane}).
From the kinetic perspective, fast motions on this hyperplane should be treated as
a  {\it relaxation} equation,
equipped with the quadratic Lyapunov function
$\delta G=\langle\!\langle\delta\cc,\delta\cc\rangle\!\rangle$,
Futhermore, we require that the linear operator of this evolution equation
 should respect Onsager's symmetry
requirements (selfadjointness with respect to the entropic scalar product).
This latter crucial requirement describes fast motions under
the frozen slow evolution in the similar way, as {\it all} the  motions near the equilibrium.

Let us consider now the manifold $\bOmega_0$ which is not the invariant
manifold of the reduced description but, by our assumption, is located
close to it. Consider a state $\cc_0\in\bOmega_0$, and the states
$\cc_0+\delta\cc$ close to it. Further, let us consider an equation
\begin{equation}
\label{RELAX_SYMM}
\dot{\delta\cc}=\LL'_{\cc_0}\delta\cc.
\end{equation}
Due to the properties of the operator $\LL'_{\cc_0}$ (\ref{SYMMETRIC}),
this equation can be regarded as a model of the
assumed true relaxation equation near the true manifold of the reduced
description. For this reason, we shall use the symmetric operator
$\LL'_{\cc}$ (\ref{SYMMETRIC}) {\it instead} of the linear operator $\LL_{\cc}$
 when constructing the corrections.

\subsection{Symmetric iteration}
Let the manifold $\bOmega_0$ and the corresponding projector
$\PP_0$ are the initial approximation to the invariant
manifold of the reduced description.
The dynamic correction $\cc_1=\cc_0+\delta\cc$
is found upon solving the following system of linear algebraic
equations:
\begin{equation}
\label{ITERATION}
\left[1-\PP_0\right]\left[\JJ(\cc_0) +
\LL'_{\cc_0}\delta\cc\right]=0,\
\PP_0\delta\cc=0.
\end{equation}
Here $\LL'_{\cc_0}$ is the matrix (\ref{SYMMETRIC})
taken in the states on the manifold $\bOmega_0$. An important
technical point here is that the linear system
(\ref{ITERATION}) always
has the unique solution for any choice of the manifold $\bOmega$.
This point is crucial since it guarantees the opportunity of carrying
out the correction process for arbitrary number of steps.

\section{The method of invariant manifold}
\label{MIM}
We shall now combine together the two procedures discussed above.
The resulting method of invariant manifold intends to seek iteratively
the reduced description, starting with an initial approximation.

(i). {\it Initialization}. In order to start the procedure, it is
required to choose the initial  manifold $\bOmega_0$,
and to derive corresponding thermodynamic projector $\PP_0$.
In the majority of cases, initial manifolds are available in two
different ways. The first case are the  quasi-equilibrium manifolds described
in the section \ref{partial_eq}.
The macroscopic parameters are $M_i=c_i=\langle\mm_i,\cc\rangle$, 
where $\mm_i$ is the unit vector corresponding to the specie $A_i$. 
The quasi-equilibrium manifold,  $\cc_0(M_1,\dots,M_k,B_1,\dots,B_l)$,
compatible with the conservation laws, is the solution to the variational
problem:
\begin{eqnarray}
\label{QE}
G\to{\rm min\ }, && \langle\mm_i,\cc\rangle=c_i, 
\ i=1,\dots,k,\\\nonumber
&&\langle\bb_j,\cc\rangle=B_j, \ j=1,\dots,l .
\end{eqnarray}
In the case of quasi--equilibrium approximation, the corresponding
thermodynamic projector can be written most straightforwardly in
terms of the variables $M_i$:
\begin{equation}
\label{PROJ-QE}
\PP_0\xx=\sum_{i=1}^{k}\frac{\partial\cc_0}{\partial c_i}
\langle\mm_i,\xx\rangle+
\sum_{i=1}^{l}\frac{\partial\cc_0}{\partial B_i}
\langle\bb_i,\xx\rangle.
\end{equation}
For quasi-equilibrium manifolds, a reparameterization with the set
(\ref{param_thermo}), (\ref{param_other}) is {\it not} necessary (\cite{GK92}; \cite{GK94}).

The second source of initial approximations are  quasi-stationary
manifolds (section \ref{QSS}).  Unlike
the quasi-equilibrium case, the quasi-stationary manifolds must be
reparameterized in order to construct the thermodynamic projector.

(ii). {\it Corrections.} Iterations are organized in accord with the rule:
If $\cc_m$ is the $m$th approximation to the invariant manifold, then
the correction $\cc_{m+1}=\cc_{m}+\delta\cc$ is found from the linear
algebraic equations,
\begin{eqnarray}
\label{corr_eq}
[1-\PP_{m}](\JJ(\cc_m)+\LL'_{\cc_m}\delta\cc)&=&0,\\
\PP_m\delta\cc&=&0.\label{corr_cond}
\end{eqnarray}
Here $\LL'_{\cc_m}$ is the symmetric matrix (\ref{SYMMETRIC})
evaluated at the $m$th approximation. The  projector
$\PP_{m}$ is not obligatory thermodynamic at that step, and it is taken as follows:
\begin{equation}
\label{proj_nonthermo}
\PP_m\xx=\sum_{i=1}^{k}\frac{\partial\cc_m}{\partial c_i}
\langle\mm_i,\xx\rangle+
\sum_{i=1}^{l}\frac{\partial\cc_m}{\partial B_i}
\langle\bb_i,\xx\rangle.
\end{equation}
(iii). {\it Dynamics.} Dynamics on the $m$th manifold is obtained with
the thermodynamic re-parameterization.

In the next section we shall illustrate how this all works.

\section{Illustration: Two-step catalytic reaction}
\label{EX}
Here we consider a two-step four-component reaction with one
catalyst $A_2$:
\begin{equation}
\label{mech_ex}
A_1+A_2\rightleftharpoons\ A_3\rightleftharpoons A_2 + A_4.
\end{equation}
We assume the Lyapunov function of the form (\ref{gfun}),
$G=\sum_{i=1}^{4}c_i[\ln(c_i/c_i^{\rm eq})-1]$.
The kinetic equation for the four--component vector of concentrations,
$\cc=(c_1,c_2,c_3,c_4)$, has the form
\begin{equation}
\label{1exequ}
\dot{\cc}=\bgamma_1W_1+\bgamma_2W_2.
\end{equation}
Here $\bgamma_{1,2}$ are stoichiometric vectors,
\begin{equation}
\label{stochio_ex}
\bgamma_1=(-1,-1,1,0),\  \bgamma_2=(0,1,-1,1),
\end{equation}
while functions $W_{1,2}$ are reaction rates:
\begin{equation}
\label{rates_ex}
W_1=k^+_1c_1c_2-k^-_1c_3,\ 
W_2=k^+_2c_3- k^-_2c_2 c_4.
\end{equation}
Here $k_{1,2}^{\pm}$ are reaction rate constants. 
The system under consideration has two conservation laws,
\begin{equation}
\label{cons_ex}
c_1+c_3+c_4=B_1,\ c_2 +c_3=B_2,
\end{equation}
or $\langle\bb_{1,2},\cc\rangle=B_{1,2}$, where
$\bb_1=(1,0,1,1)$ and $\bb_2=(0,1,1,0)$. The nonlinear system
(\ref{1exequ}) is effectively two-dimensional, and we consider a
one-dimensional reduced description. 

We have chosen the concentration of the specie
$A_1$ as the variable of reduced description:
$M=c_1$, and 
$c_1=\langle \mm,\cc\rangle$, where $\mm=(1,0,0,0)$.
The initial manifold $\cc_0(M)$ was taken as the quasi-equilibrium
approximation, i.e. the vector function $\cc_0$ is the solution to
the problem:
\begin{equation}
\label{qe}
G\to{\rm min}\ {\rm for}\
\langle\mm,\cc\rangle=c_1,\ \langle\bb_1,\cc\rangle=B_1,\ \langle\bb_2,\cc\rangle=B_2.
\end{equation}
The solution to the problem (\ref{qe}) reads:
\begin{eqnarray}
\label{1ex1}
c_{01}&=&c_1,\\\nonumber
c_{02}&=&B_2-\phi(c_1), \\\nonumber
c_{03}&=&\phi(c_1),\\\nonumber
c_{04}&=&B_1-c_1-\phi(c_1),\\\nonumber
\phi(M)&=&A(c_1)-\sqrt{A^2(c_1)-B_2(B_1-c_1)},\\\nonumber
A(c_1)&=&\frac{B_2(B_1-c^{\rm eq}_1)+c_3^{\rm eq}(c_1^{\rm eq}+
c_3^{\rm eq}-c_1)}{2c_3^{\rm eq}} .
\end{eqnarray}
The thermodynamic projector associated with the manifold
(\ref{1ex1}) reads:
\begin{equation}
\label{proj_0}
\PP_0\xx=\frac{\partial\cc_0}{\partial c_1}\langle\mm,\xx\rangle
+\frac{\partial\cc_0}{\partial B_1}\langle\bb_1,\xx\rangle+
\frac{\partial\cc_0}{\partial B_2}\langle\bb_2,\xx\rangle.
\end{equation}
Computing
$\bDelta_0=[1-\PP_0]\JJ(\cc_0)$
we find that the inequality (\ref{DEFECT}) takes place, and thus
the manifold $\cc_0$ is not invariant. The first correction, $\cc_1=\cc_0+\delta\cc$,
is found from the linear algebraic system (\ref{corr_eq})
\begin{eqnarray}
\label{LIN_SYSTEM}
(1-\PP_0)\LL'_{0}\delta\cc&=&-[1-\PP_0]\JJ(\cc_0),\\\nonumber
\delta c_1&=&0 \nonumber \\
\delta c_1+\delta c_3+\delta c_4&=&0 \nonumber \\
\delta c_3+\delta c_2&=&0,
\label{correction1}
\end{eqnarray}
where the symmetric $4\times 4$ matrix $\LL'_{0}$ has the form
(we write $0$ instead of $\cc_0$ in the subscript in order to simplify
notations):
\begin{equation}
\label{SL}
L'_{0, kl}=-\gamma_{1k}
\frac{W_1^+(\cc_0)+W_1^-(\cc_0)}{2}
\frac{\gamma_{1l}}{c_{0l}}
-
\gamma_{2k}
\frac{W_2^+(\cc_0)+W_2^-(\cc_0)}{2}
\frac{\gamma_{2l}}{c_{0l}}
\end{equation}
The explicit solution $\cc_1(c_1,B_1,B_2)$ to the linear system
(\ref{LIN_SYSTEM}) is easily
found, and we do not reproduce it here. 
The process was iterated. On the $k+1$ iteration, the following projector
$\PP_{k}$ was used:
\begin{equation}
\label{proj_k}
\PP_k\xx=\frac{\partial\cc_k}{\partial c_1}\langle\mm,\xx\rangle+
\frac{\partial\cc_k}{\partial B_1}\langle\bb_1,\xx\rangle+
\frac{\partial\cc_k}{\partial B_2}\langle\bb_2,\xx\rangle.
\end{equation}
Notice that projector $\PP_k$ (\ref{proj_k}) is the thermodynamic
projector only if $k=0$. As we have already mentioned it above,
in the process of finding the corrections to the manifold, 
the non-thermodynamic projectors are allowed.
The linear equation at the $k+1$ iteration is thus
obtained by replacing $\cc_0$, $\PP_0$, and $\LL'_0$ with
$\cc_k$, $\PP_k$, and $\LL'_k$ in all the entries of the Eqs.\ 
(\ref{LIN_SYSTEM}) and (\ref{SL}).

Once the manifold $\cc_k$ was obtained on the $k$th iteration,
we derived the corresponding dynamics by introducing the thermodynamic
parameterization (and the corresponding thermodynamic projector)
with the help of the function (\ref{param_thermo}).
The resulting dynamic equation for the variable $c_1$
in the $k$th approximation has the form:
\begin{equation}
\label{1ex3}
\langle\bnabla G\bigm|_{\cc_k},\partial\cc_k/\partial c_1\rangle\dot{c_1}=
\langle\bnabla G\bigm|_{\cc_k},\JJ(\cc_k)\rangle.
\end{equation}
Here $[\bnabla G\bigm|_{\cc_k}]_i=\ln[c_{ki}/ c^{\rm eq}_i]$.

Analytic results were compared with the results of the numerical
integration. The following set of parameters was used:
\begin{eqnarray*}
k^+_1=1.0,\ k^-_1=0.5,\  k^+_2=0.4 ,\ k^-_2=1.0;\\
c_1^{\rm eq}=0.5,\ c_2^{\rm eq}=0.1,\ c_3^{\rm eq}=0.1,\ c_4^{\rm eq}=0.4,\\
B_1=1.0    ,\ \ B_2=0.2   .
\end{eqnarray*}
Direct numerical integration of the system has demonstrated
that the manifold $c_3= c^{\rm eq}_3$ in the plane $(c_1,c_3)$
attracts all individual trajectories. Thus, the reduced description in this
example should extract this manifold.

Fig.\ \ref{Fig1}  demonstrates the quasi--equilibrium manifold
(\ref{1ex1}) and first two corrections found analytically. It is apparent that
while the initial quasi-equilibrium
approximation is in a poor agreement with the reduced description, the
corrections {\it rapidly} improve the situation. This confirms our expectation
of an advantage of using iteration methods in comparison to methods
based on a small parameter expansions.

\section{Method of invariant manifold without a priori parameterization}
\label{PARAMETERIZATION}
Formally, the method of invariant manifold does not require a global parameterization of the 
manifolds. However, in most of the cases, one makes use of a priori defined 
``macroscopic'' variables
$M$. This is motivated by the choice of quasi-equilibrium initial approximations.

Let a manifold $\bOmega$ be defined in the phase space of the system, its tangent
space in the point $\cc$  be $T_{\cc}\bOmega$. How to define the projector
of the whole concentrations space onto $T_{\cc}\bOmega$ without using any 
a priori parameterization of $\bOmega$? 

The basis of the answer to this question is the condition of thermodynamicity 
(\ref{therm1}). Let us denote $E$ as the concentration space, and consider the
problem of the choice of the projector in the quadratic approximation
to the thermodynamic potential $G$:

\begin{equation}
G_{\rm q}=\langle \bg,\HH_{\cc}\Delta\cc\rangle
+\frac{1}{2}\langle\Delta\cc,\HH_{\cc}\Delta\cc\rangle
=\langle\!\langle\bg,\Delta\cc\rangle\!\rangle+
\frac{1}{2}\langle\!\langle\Delta\cc,\Delta\cc\rangle\!\rangle,
\label{Gquadratic}
\end{equation}
where $\HH_{\cc}$ is the matrix of the second-order derivatives of $G$ (\ref{MATRIX}),
$\bg=\HH^{-1}_{\cc}\bnabla G$, $\Delta\cc$ is the deviation of the concentration
vector from the expansion point. 

Let a linear subspace $T$ be given in the concentrations space $E$.
{\it Problem:} For every $\Delta\cc+T$, and for every $\bg\in E$, define  a subspace
$L_{\Delta\cc}$ such that:
(i) $L_{\Delta\cc}$ is a complement of $T$ in $E$: 
\[L_{\Delta\cc}+T=E,\ L_{\Delta\cc}\cap T=\{\bZERO\}.\]
(ii) $\Delta\cc$ is the point of minimum of
$G_{\rm q}$ on $L_{\Delta\cc}+\Delta\cc$:
\begin{equation}
\label{minQ}
\Delta\cc=\arg\min_{\xx-\Delta\cc\in L_{\Delta\cc}}G_{\rm q}(\xx).
\end{equation}
Besides (i) and (ii), we also impose the requirement of
a {\it maximal smoothness} (analyticity) on $L_{\Delta\cc}$ as a function of
$\bg$ and $\Delta\cc$. Requirement (\ref{minQ}) implies that $\Delta\cc$ is the quasi-equilibrium
point for the given $L_{\Delta\cc}$, while the problem in a whole is the
{\it inverse} quasi-equilibrium problem: We construct $L_{\Delta\cc}$ such that
$T$ will be the quasi-equilibrium manifold. Then
subspaces $L_{\Delta\cc}$ will actually be the kernels of the quasi-equilibrium projector.

Let $\ff_1,\dots,\ff_k$ be the orthonormalized with respect to 
$\langle\!\langle\cdot,\cdot\rangle\!\rangle$ scalar product basis of $T$,
vector $\hh$ be orthogonal to $T$, $\langle\!\langle\hh,\hh\rangle\!\rangle=1$,
$\bg=\alpha\ff_1+\beta\hh$. Condition (\ref{minQ})
implies that the vector $\bnabla G$ is orthogonal to
$L_{\Delta\cc}$ in the point $\Delta\cc$.

Let us first consider the case $\beta=0$. The requirement of analyticity
of $L_{\Delta\cc}$ as the function of $\alpha$ and $\Delta\cc$ implies
$L_{\Delta\cc}=L_{\bZERO}+o(1)$, where
$L_{\bZERO}=T^{\perp}$ is the orthogonal completement of $T$ with respect
to scalar product $\langle\langle\cdot,\cdot\rangle\rangle$.
The constant solution, $L_{\Delta\cc}\equiv L_{\bZERO}$  also
satisfies (\ref{minQ}). Let us fix $\alpha\ne0$, and extend this latter solution
to $\beta\ne0$. With this, we obtain a basis, $\bl_1,\dots,\bl_{n-k}$.
Here is the simplest construction of this basis:
\begin{equation}
\label{ort1}
\bl_1=\frac{\beta\ff_1-(\alpha+\Delta c_1)\hh}{(\beta^2+(\alpha+\Delta c_1)^2)^{1/2}},
\end{equation}
where $\Delta c_1=\langle\langle\Delta\cc,\ff_1\rangle\rangle$
is the first component in the expansion, $\Delta\cc=\sum_i\Delta c_i\ff_i$.
The rest of the basis elements, $\bl_2,\dots,\bl_{n-k}$ form
the orthogonal completement of $T\oplus(\hh)$ with respect to scalar
product  $\langle\langle\cdot,\cdot\rangle\rangle$, $(\hh)$ is the line spanned by
$\hh$. 

Dependence $L_{\Delta\cc}$ (\ref{ort1}) on $\Delta\cc$, $\alpha$ and $\beta$
is singular: At $\alpha+\Delta c_1$, vector $\bl_1\in T$, and then  $L_{\Delta\cc}$
is not the completement of $T$ in $E$ anymore.
For $\alpha\ne0$, dependence $L_{\Delta\cc}$ gives one of the solutions to the inverse 
quasi-equilibrium problem in the neighborhood of zero in $T$. We are interested only in the limit,
\begin{equation}
\label{lim1}
\lim_{\Delta\cc\to\bZERO}L_{\Delta\cc}={\rm Lin}\left\{
\frac{\beta\ff_1-\alpha\hh}{\sqrt{\alpha^2+\beta^2}},\bl_2,\dots,\bl_{n-k}\right\}.
\end{equation}

Finally, let us define now the projector $\PP_{\cc}$ of the space $E$ onto $T_{\cc}\bOmega$.
If $\HH^{-1}_{\cc}\bnabla G\in T_{\cc}\bOmega$,
then $\PP_{\cc}$ is the orthogonal projector with respect
to the scalar product
 $\langle\langle\cdot,\cdot\rangle\rangle$:

\begin{equation}
\PP_{\cc}\zz=\sum_{i=1}^k\ff_i\langle\langle \ff_i,\zz\rangle\rangle.
\end{equation}
If $\HH^{-1}_{\cc}\bnabla G \notin T_{\cc}\bOmega$, then, according to
Eq.\ (\ref{lim1}),
\begin{equation}
\label{result_projector}
\PP_{\cc}\zz=\frac{\langle\langle \ff_1,\zz\rangle\rangle
-\langle\langle \bl_1,\zz\rangle\rangle
\langle\langle \ff_1,\bl_1\rangle\rangle}
{1-\langle\langle \ff_1,\bl_1\rangle\rangle^2}\ff_1+
\sum_{i=2}^k\ff_i\langle\langle \ff_i,\zz\rangle\rangle,
\end{equation}
where 
$\{\ff_1,\dots,\ff_k\}$ is the orthonormal with respect to
$\langle\langle\cdot,\cdot \rangle\rangle$ basis of $T_{\cc}\bOmega$,
$\hh$ is orthogonal to $T$, $\langle\langle \hh,\hh\rangle\rangle=1$,
$\HH^{-1}_{\cc}\bnabla G=\alpha\ff_1+\beta\hh$, 
$\bl_1=(\beta\ff_1-\alpha\hh)/\sqrt{\alpha^2+\beta^2}$,
$\langle\langle \ff_1,\bl_1\rangle\rangle=\beta/\sqrt{\alpha^2+\beta^2}$.

Thus, for solving the invariance equation iteratively, one needs only projector $\PP_{\cc}$
(\ref{result_projector}), and one does not need a priori parameterization of $\bOmega$
anymore.

\section{Method of invariant grids}
\label{grid}

Grid-based approximations of manifolds are attractive
from the computational perspective. Since no a priori parameterization
is required in the method of invariant manifold, in this section
we develop its grid-based realization.
Let us consider a regular grid $Q$ in $R^k$, and its
mapping $F$ into the concentrations space $E$.
It makes sense to consider only
$F$ which map a finite part of the grid into
the phase space $V$. This part of the map is termed {\it essential}.
Extension of the map $F$ onto the rest of the nodes is done by a
simple (for example, linear) extrapolation of the essential part
(in practice, one needs to extrapolate only onto the next
neighbors of the essential nodes).

Let operators of grid differentiation $D_i$ 
where be defined for functions on the grid, where $i=1,\dots,k$
label grid coordinates $x_i$. 
With this, the tangent space to the image of the grid in the
point $\cc(x)=F(x)$ is defined for each node of the grid $x$:
\begin{eqnarray}
T_x={\rm Lin}\{\varphi_1,\dots,\varphi_k\},\nonumber\\
\varphi_i=D_i\cc(x)=(D_ic_1(x),\dots,D_ic_n(x)).
\end{eqnarray}
The grid is termed invariant if, for each essential node,
\[ \JJ(\cc(x))\in T_x.\]
For the essential nodes, we write down the invariance equation
with the projector, $P_{\cc(x)}: E\to T_x$:
This equation is solved using the Newton method as it was described
above in the section \ref{MIM}. A good initial approximation
is a linear map of the grid on the affine manifold corresponding 
to slow relaxation in the vicinity of the equilibrium.
It is convenient to take this map isometric with respect
to the metrics generated by the entropic scalar product
in the equilibrium.

If the vector field of the reduced model, 
$\dot{\cc}=\PP_{\cc(x)}\JJ(\cc(x))$,
is defined on the nodes $F(x)$, then one can define the
dynamics $\dot{x}_i$ on the nodes. In order to do this,
we expand $\dot{\cc}$ over $\varphi_i$: 
$\dot{\cc}=\sum_{i=1}^ka_i\varphi_i$.
The dynamics on the nodes is then defined by equations, 
$\dot{x}_i=a_i$.
Using interpolation, we can define the vector field
$\dot{x}$ within the essential cells of the grid (those cells for
which the all the nodes are essential). The system of equations thus
obtained models the dynamics on the invariant manifold.

The essence of this construction is that, by solving a set
of uncomplicated linear equations arising from linearization
of the invariance equations on the nodes one gets a reliable
numerical scheme for constructing invariant manifolds. 
The use of the grid differentiation rather than a differentiable
approximation to the manifold makes the scheme suited for
parallel realizations. We stress it once again that
such realizations are only possible if no a priori global
parameterization of manifolds is required. 
Further refinements of the scheme,
taking into account the process of moving the inessential nodes
into the phase space, and the opposite process of essential
nodes leaving the phase space can be done based in the same way
as for grid-based data analysis \citep{GR99,GZ01}.

\section{Method of invariant manifold for open systems}
\label{open}
One of the problems to be focused on when studying closed
systems is to prepare extensions of the result for
open or driven by flows systems.
External flows are usually taken into account
by addidinal terms in the kinetic equations (\ref{reaction}):
\begin{equation}
\label{external}
\dot{\cc}=\JJ(\cc)+\bPi.
\end{equation}
{\it Zero-order approximation} assumes that the flow does
not change the invariant manifold.
Equations of the reduced dynamics, however, do change:
Instead of $\JJ(\cc(M))$ we substitute
$\JJ(\cc(M))+\bPi$ into Eq.\ (\ref{dyn}):
\begin{equation}
\label{zero}
\dot{M}_i=\sum_{j=1}^pN^{-1}_{ij}\langle \bnabla M_j\big|_{\cc(M)},
\JJ(\cc(M))+\bPi\rangle.
\end{equation}
Zero-order approximation assumes that the fast dynamics in the
closed system strongly couples the variables $\cc$, so that
flows cannot influence this coupling.

{\it First-order approximation} takes into account the shift
of the invariant manifold by $\delta\cc$. Equations for Newton's iterations have
the same form (\ref{ITERATION}) but instead of the vector field
$\JJ$ they take into account the presence of the flow:
\begin{equation}
\label{one}
[1-\PP_{\cc}](\bPi+\LL'_{\cc}\delta\cc)=0,\ \PP_{\cc}\delta\cc=0,
\end{equation}
where projector $\PP_{\cc}$ corresponds to the unperturbed manifold.

The first-order approximation means that fluxes change the coupling
between the variables (concentrations). It is assumed that
these new coupling is also set instantaneously (neglect of inertia).

{\it Remark.} Various realizations of the first-order approximation 
in physical and chemical dynamics implement the viewpoint
of an infinitely small chemical reactor driven by the flow.
In other words, this approximation is applicable
in the Lagrangian system of coordinates \citep{KGDN98,ZKD00}. 
Transition to Eulerian coordinates is possible
but the relations between concentrations and the flow will change
its form. In a contrast, the more simplistic zero-order approximation
is equally applicable in both the coordinate system, if it is valid.

\section{Conclusion}
\label{conclusion}

In this paper, we have
presented the method for constructing the invariant manifolds
for reducing systems of chemical kinetics.
Our approach to computations of invariant manifolds
of dissipative systems is close in spritit to the
Kolmogorov-Arnold-Moser theory of invariant tori of Hamiltonian systems
\citep{Arnold63,Arnold83}: We also base our consideration on the Newton
method instead of Taylor series expansions \citep{Beyn98}, and systematically
use duality structures.
Recently, a version of an approach based on the invarinace equations
was rediscovered by \cite{Kazantzis00}. He was solving  the invariance 
equation by a Taylor series expansion. A counterpart of Taylor series expansions
for constructing the slow invariant manifolds in the classical kinetic theory
is the famous Chapman-Enskog method. The question of how this 
compares to iteration methods was studied extensively for 
certain classes of Grad moment equations \citep{GK96a,KDN97a,K00}.

The thermodynamic parameterization and the selfadjoint linearization arise
in a natural way in the problem of finding slowest invariant manifolds for
closed systems. This also leads to  various applications
in different approaches to reducing the description, in particular, to
a  thermodynamically consistent version
of the intrinsic low-dimensional manifold, and to  model kinetic equations for lifting
the reduced dynamics. Use of the thermodynamic projector
makes it unnecessary global parameterizations of manifolds, and
thus leads to computationally promising grid-based realizations.

Invariant manifolds are constructed for closed space-independent chemical systems.
We also describe how to use these manifolds for modeling open and distributed
systems.

\newpage

\begin{figure}


\caption{Images of the initial quasi-equilibrium manifold (bold line)
 and the first two corrections
(solid normal lines) in the phase plane
$[c_1,c_3]$ for two-step catalytic reaction (\ref{mech_ex}).
Dashed lines are individual trajectories.}
\label{Fig1}
\end{figure}


\begin{thebibliography}{}



\bibitem[Ansumali \& Karlin(2000)]{AK00} Ansumali, S., \&  Karlin, I.\ V. (2000). Stabilization of
the Lattice Boltzmann method by the $H$ theorem: A numerical test.
{\it Phys.\ Rev.\ E,} {\bf 62(6)}, 7999-8003.


\bibitem[Ansumali \& Karlin(2002)]{AK02}Ansumali S.,\ \&  Karlin, I.\ V. (2002). Single relaxation time
model for entropic Lattice Boltzmann methods.  {\it Phys.\ Rev.\ E,} {\bf 65} 056312(1-9).

\bibitem[Ansumali \& Karlin(2002a)]{AK02a}
 Ansumali, S., \&  Karlin, I.V. (2002a). Entropy function approach
to the lattice Boltzmann method. {\it J.\ Stat.\ Phys.,} {\bf 107(1/2)} 291-
308.

\bibitem[Arnold(1963)]{Arnold63}
Arnold, V.\ I.\ (1963). Proof of a theorem of A.\ N.\ Kolmogorov on the 
invariance of quasi-periodic motions under small perturbations of the 
Hamiltonian. (English translation). {\it Russian Mathematical Surveys,} {\bf 18}, 9-36.

\bibitem[Arnold(1983)]{Arnold83}Arnold, V.\ I.\ (1983). 
{\it Geometrical methods in the theory of ordinary  differential equations.}
New York: Springer.

\bibitem[Beyn \& Kless(1998)]{Beyn98}
Beyn, W.-J., \& Kless, W. (1998). Numerical Taylor expansions of invariant 
manifolds in large dynamical systems. {\it Numerische Mathematik,} {\bf 80}, 1-38.

\bibitem[Bhatnagar, Gross \& Krook(1954)]{BGK}Bhatnagar, P.\ L., 
Gross, E.\ P., \&  Krook, M. (1954).
A model for collision processes in gases. 
Small amplitude processes in charged and neutral one-component systems.
{\it Phys.\ Rev.,}
{\bf 94}, 511-525.


\bibitem[Bobylev(1988)]{Bobylev88}Bobylev, A.\ V.\ (1988).  
The theory of the nonlinear spatially uniform Boltzmann equation for Maxwell molecules. 
{\it Mathematical physics reviews,} {\bf 7}, 111-233.

\bibitem[Bowen, Acrivos \&  Oppenheim(1963)]{Bowen63}
Bowen, J.\  R., Acrivos, A., \& Oppenheim, A.\  K. (1963). Singular Perturbation 
Refinement to Quasi-Steady State Approximation in Chemical Kinetics, 
{\it Chemical Engineering Science,} {\bf 18}, 177-188.

\bibitem[Bykov, Gorban \& Yablonskii(1982)]{BGY82}
Bykov, V.\ I., Gorban, A.\ N., \& Yablonskii, G.\ S. (1982). Description of 
nonisothermal reactions in terms of Marcelin-de Donder kinetics and its 
generalizations. {\it React.\  Kinet.\ Catal.\  Lett.,} {\bf 20}, 261-265.

\bibitem[Bykov, Yablonskii \& Akramov(1977)]{BYA77}
Bykov, V.\ I.,  Yablonskii, G.\ S., \& Akramov, T.\ A. (1977). 
The rate of the free energy decrease in the course of the complex chemical reaction.
{\it Dokl.\ Akad.\ Nauk  USSR,} {\bf 234 (3)},  621-634.

\bibitem[Chen(1988)]{Chen88}Chen, J.-Y. (1988). A general procedure for constructing 
reduced reaction 
mechanisms with given independent relations. {\it Combustion Science and 
Technology,} {\bf 57}, 89-94.

\bibitem[Chicone(2000)]{Chicone00}
Chicone, C., \& Swanson, R. (2000). Linearization via the Lie derivative. 
{\it Electron.\ J.\  Diff.\ Eqns.,} Monograph, 02, 
http://ejde.math.swt.edu or http://ejde.math.unt.edu
ftp ejde.math.swt.edu or ejde.math.unt.edu (login: ftp)

\bibitem[Dimitrov(1982)]{Dimitrov82}Dimitrov, V.I. (1982). 
{\it Prostaya kinetika [Simple Kinetics]}. Novosibirsk: Nauka.


\bibitem[De Donder \& Van Rysselberghe(1936)]{DeDonder36}
De Donder, T., \& Van Rysselberghe, P. (1936). {\it Thermodynamic theory of 
affnity. A book of principles.} Stanford: University Press.

\bibitem[Dukek, Karlin \& Nonnenmacher(1997)]{DKN97}
Dukek, G., Karlin, I.\ V., \& Nonnenmacher, T.\ F. (1997). Dissipative brackets 
as a tool for kinetic modeling. {\it Physica A}, {\bf 239(4)}, 493-508.

\bibitem[Feinberg(1972)]{Feinberg72}Feinberg, M. (1972). 
On chemical kinetics of a certain class. {\it Arch.\ Rational 
Mech. Anal.,} {\bf 46(1)}, 1-41.

\bibitem[Fraser(1988)]{Fraser88}Fraser, S.\ J. (1988). 
The steady state and equilibrium approximations: A 
geometrical picture. {\it J.\ Chem.\ Phys.}, {\bf 88(8)},  4732-4738.

\bibitem[Gear(1971)]{Gear71}Gear, C.\ W. (1971).
 {\it Numerical initial value problems in ordinary differential equations.} 
Prentice-Hall, Englewood Cliffs, NJ.

\bibitem[Gorban(1984)]{G84}
Gorban, A.N. (1984). {\it Obkhod ravnovesiya [Equilibrium encircling].} 
 Novosibirsk: Nauka.


\bibitem[Gorban \& Karlin(1992)]{GK92}Gorban, A.\ N., \& Karlin, I.\ V. (1992). 
Thermodynamic parameterization. {\it Physica A}, {\bf 190}, 393-404 .

\bibitem[Gorban \& Karlin(1992a)]{GK92a} Gorban, A.\ N., \& Karlin, I.\ V. (1992a). 
The constructing of 
invariant manifolds for the Boltzmann equation, {\it Adv.\ Model. and 
Analysis C}, {\bf 33(3)}, 39-54.

\bibitem[Gorban \& Karlin(1992b)]{GK92b}
Gorban, A.\ N., \& Karlin, I.\ V. (1992b). Coarse-grained quasi-
equilibrium approximations for kinetic equations. {\it Adv.\ Model. and 
Analysis C}, {\bf 35(1)}, 17-27.

\bibitem[Gorban \& Karlin(1992c)]{GK92c}Gorban, A.\ N., \& Karlin, I.\ V. (1992c). 
H-theorem for generalized 
models of the Boltzmann equation. {\it Adv.\ Model. and Analysis C}, {\bf 
33(3)}, 33-38.

\bibitem[Gorban \& Karlin(1994)]{GK94}Gorban, A.\ N., \& Karlin, I.\ V. (1994). 
Method of invariant 
manifolds and regularization of acoustic spectra. {\it Transport Theory and 
Stat.\ Phys.}, {\bf 23}, 559-632.

\bibitem[Gorban \& Karlin(1994a)]{GK94a} 
Gorban, A.\ N., \& Karlin, I.\ V. (1994a). General approach to 
constructing models of the Boltzmann equation. {\it Physica A}, {\bf 206}, 
401-420.

\bibitem[Gorban \& Karlin(1996)]{GK96}Gorban, A.\ N., \& Karlin, I.\ V., (1996). 
Scattering rates versus moments: Alternative Grad equations. {\it Phys.\ Rev. E},
 {\bf 54(4)}, R3109-R3113.

\bibitem[Gorban \& Karlin(1996a)]{GK96a}Gorban, A.\ N., \&  Karlin, I.\ V., (1996a)
Short-wave limit
of hydrodynamics: A soluble example. {\it Phys.\ Rev.\ Lett.,} {\bf
77}, 282-285. 

\bibitem[Gorban, Karlin, Zmievskii \& Nonnenmacher(1996)]{GKZN96}
Gorban, A.\ N., Karlin, I.\ V., Zmievskii, V.\ B., \& Nonnenmacher, T.\ F. 
(1996). Relaxational trajectories: global approximations. {\it Physica A}, 
{\bf 231}, 648-672. 

\bibitem[Gorban, Karlin \& Zmievskii(1999)]{GKZ99}
Gorban, A.\ N., Karlin, I.\ V., \& Zmievskii, V.\ B. (1999). Two-
step approximation of space-independent relaxation. {\it Transp.\ Theory 
Stat.\ Phys.}, {\bf 28(3)}, 271-296.

\bibitem[Gorban, Karlin, Zmievskii \& Dymova(2000)]{GKZD00}
Gorban, A.\ N., Karlin, I.\ V., Zmievskii, V.\ B., \& Dymova S.\ V. 
(2000). Reduced description in reaction kinetics.
{\it Physica A}, {\bf 275(3-4)}, 361-379.

\bibitem[Gorban, Karlin, Ilg \& \"Ottinger(2001)]{GKIOe01}
Gorban, A.\ N., Karlin, I.\ V., Ilg, P., \& \"{O}ttinger, H.\ C. (2001). 
Corrections and enhancements of quasi-equilibrium states. {\it J.\ Non-
Newtonian Fluid Mech.}, {\bf 96(1-2)}, 203-219.


\bibitem[Gorban \& Rossiev(1999)]{GR99}
Gorban, A.\ N., \& Rossiev, A.\ A. (1999). Neural network iterative method of 
principal curves for data with gaps. {\it Journal of Computer and System 
Sciences Intrnational}, {\bf 38(5)}, 825-831.

\bibitem[Gorban \& Zinovyev(2001)]{GZ01}
Gorban, A.\ N., Zinovyev, A.\ Yu. (2001). Visualization of data by method of 
elastic maps and its applications in genomics, economics and sociology. 
Institut des Hautes Etudes Scientifiques, Preprint. IHES M/01/36. Online-
version: http://www.ihes.fr/PREPRINTS/M01/Resu/resu-M01-36.html. 


\bibitem[Grmela, Karlin \& Zmievski(2002)]{GKZ02}
 Grmela, M., Karlin, I.\ V., \& Zmievski, V.\ B. (2002). Boundary 
layer minimum entropy principles: A case study.
 {\it Phys.\ Rev.\ E} (in press).

\bibitem[Hartman(1963)]{Hartman63}
Hartman, P. (1963). On the local linearization of differential equations. {\it  Proc. 
Amer.\ Math.\ Soc.}, {\bf 14}, 568-573.

\bibitem[Hartman(1982)]{Hartman82}Hartman, P. (1982). 
{\it Ordinary differential equations}. Boston:  Birkh\"auser.

\bibitem[Karlin(1989)]{K89} Karlin, I.\ V. (1989). On the relaxation of the chemical reaction 
rate. In: {\it Mathematical Problems of Chemical Kinetics}, eds. K.\ I.\ 
Zamaraev and G.\ S.\ Yablonskii, (Nauka, Novosibirsk), 7-42. 

\bibitem[Karlin(1993)]{K93} Karlin, I.\ V. (1993). The problem of reduced description in 
kinetic theory of chemically reacting gas. {\it Modeling, Measurement and 
Control C}, {\bf 34(4)}, 1-34.

\bibitem[Karlin(2000)]{K00} Karlin, I.\ V. (2000). 
Exact summation of the Chapman-Enskog expansion
from moment equations. {\it J.\ Phys.\ A: Math.\ Gen.}, 
{\bf 33}, 8037-8046.

\bibitem[Karlin, Dukek \& Nonnenmacher(1997)]{KDN97}Karlin, I.\ V., Dukek, G., 
Nonnenmacher, T.\ F. (1997). Invariance principle 
for extension of hydrodynamics: Nonlinear viscosity. {\it Phys.\ Rev.\ E}, 
{\bf 55(2)}, 1573-1576.

\bibitem[Karlin, Dukek \& Nonnenmacher(1997a)]{KDN97a}
Karlin, I.\ V.,  Dukek, G., \&  Nonnenmacher, T.\ F. (1997a)
Gradient expansions in kinetic theory of phonons. 
{\it Phys.\ Rev.\ B,} {\bf 55}, 6324-6329.

\bibitem[Karlin, Gorban, Dukek \& Nonnenmacher(1998)]{KGDN98} 
Karlin, I.\ V., Gorban, A.\ N., Dukek, \& G., Nonnenmacher, T.\ F. 
(1998). Dynamic correction to moment approximations. {\it Phys.\ Rev.\ E}, 
{\bf 57}, 1668--1672.



\bibitem[Kato(1976)]{Kato76}Kato, T. (1976). 
{\it Perturbation theory for Linear operators.}  Berlin: Springer.

\bibitem[Kazantzis(2000)]{Kazantzis00}Kazantzis N, (2000) Singular PDEs and the problem of finding invariant manifolds for nonlinear dynamical systems. {\it Physics Letters}, {\bf A272(4)},
257-263.

\bibitem[Khibnik, Kuznetsov, Levitin \& Nikolaev(1993)]
{Khibnik93}Khibnik, A., Kuznetsov, Y., Levitin, V., \& Nikolaev, E. (1993). Continuation 
techniques and interactive software for bifurcation analysis of ODEs and 
iterated maps. {\it Physica D}, {\bf 62}, 360-370.


\bibitem[Lam \& Goussis(1994)]{Lam94}
Lam, S.H., Goussis, D. A. (1994). The CSP Method for Simplifying Ki-
netics. {\it International Journal of Chemical Kinetics}, {\bf 26}, 461-486.

\bibitem[Levenspiel(1999)]{Levenspiel99}
Levenspiel, O. (1999). Chemical Reaction Engineering. {\it Ind.\ Eng.\ Chem.\ Res.}, 
{\bf 38}, 4139-4143.

\bibitem[Levenspiel(2000)]{Levenspiel00}
Levenspiel, O. (2000). Response to Professor Yablonsky. {\it  Ind.\  Eng.\ Chem.\ Res.}, 
{\bf 39}, 3120.

\bibitem[Li, Rabitz \& T\'oth(1994)]{Li94}Li, G., Rabitz, H., \& T\'oth, J. (1994). 
A general analysis of exact nonlinear 
lumping in chemical kinetics. {\it Chem.\ Eng.\ Sci.},  {\bf 49(3)}, 343-361. 

\bibitem[Maas \& Pope(1992)]{Maas92}Maas, U., \& Pope, S.B. (1992). 
Simplifying chemical kinetics: intrinsic low-
dimensional manifolds in composition space. {\it Combustion and Flame},  {\bf 88}, 
239-264. 

\bibitem[Orlov \& Rozonoer(1984)]{Orlov84}Orlov, N. N., \& Rozonoer, L. I. (1984). 
The macrodynamics of open systems 
and the variational principle of the local potential.  {\it J. Franklin Inst.},  {\bf 318}, 283-
314 and 315-347. 

\bibitem[Rabitz(1987)]{Rabitz87}Rabitz, H. (1987). 
Chemical Dynamics and Kinetics Phenomena as revealed 
by Sensitivity Analysis Techniques. {\it Chem.\ Rev.},  {\bf 87}, 101-112.

\bibitem[Rabitz, Kramer \& Dacol(1983)]{Rabitz83}Rabitz, H., Kramer, M., \& Dacol, D. (1983). 
Sensitivity analysis in chemical 
kinetics. {\it Ann.\ Rev.\ Phys.\ Chem.},  {\bf 34}, 419-461.

\bibitem[Roose \& Spence(1990)]{Roose90}Roose, De Dier, B., \& Spence, A., eds. (1990). 
{\it Continuation and bifurcations 
numerical techniques and applications}. Dordrecht: Kluwer.

\bibitem[Roussel \& Fraser(1990)]{Roussel90}Roussel, M.R., \& Fraser, S.J. (1990). 
Geometry of the steady-state 
approximation: Perturbation and accelerated convergence methods.  {\it J.\  Chem.\ 
Phys.}, {\bf 93}, 1072-1081.

\bibitem[Roussel \& Fraser(1991)]{Roussel91}Roussel, M.R., \& Fraser, S.J. (1991). 
On the geometry of transient relaxation. 
{\it J.\ Chem.\ Phys.}, {\bf  94}, 7106-711.

\bibitem[Segel \& Slemrod(1989)]{Segel89}Segel, L.A., \& Slemrod, M. (1989). 
The quasi-steady-state assumption: A case 
study in perturbation. {\it SIAM Rev.,} {\bf 31}, 446-477.

\bibitem[Struchtrup \& Weiss(1998)]{Struchtrup98}
Struchtrup, H., \& Weiss, W. (1998). Maximum of the local entropy production becomes 
minimal in stationary processes. {\it Phys.\ Rev.\ Lett.,} {\bf 80}, 5048-5051.

\bibitem[Strygin \& Sobolev(1988)]{Strygin88}
Strygin V.\ V., \& Sobolev, V.\ A. (1988). {\it 
Spliting of motion by means of integral manifolds}. Moscow: Nauka.

\bibitem[T\'oth, Li, Rabitz \& Tomlin(1997)]{Toth97}
T\'oth, J., Li, G., Rabitz, H., \& Tomlin, A. S. (1997). The effect of lumping and 
expanding on kinetic differential equations. {\it SIAM J. Appl. Math.,} {\bf 57(6)}, 
1531-1556. 

\bibitem[Vasil'eva, Butuzov \& Kalachev(1995)]{Vasil'eva95}
Vasil'eva A.B., Butuzov V.F., \& Kalachev L.V. (1995). {\it  The boundary 
function method for singular perturbation problems}, Philadelphia: SIAM.

\bibitem[Volpert \& Hudjaev(1985)]{Volpert85}Volpert, A. I., \& Hudjaev, S. I. (1985).
{\it  Analysis in classes of discontinuous 
functions and the equations of mathematical physics}. Dordrecht: Nijhoff.


\bibitem[Wei \& Prater(1962)]{Wei62}Wei, J., \&  Prater, C. (1962). 
The structure and analysis of complex reaction 
systems. {\it  Adv.\ Catalysis,} {\bf 13}, 203-393. 

\bibitem[Yablonsky(2000)]{Y00}Yablonsky, G.\ S. (2000). 
Comments on a commentary by Professor 
Levenspiel, {\it Ind.\ Eng.\ Chem.\ Res.}, {\bf 39}, 3120.

\bibitem[Yablonskii, Bykov, Gorban \& Elokhin(1991)]{YBGE91}
Yablonskii, G.S., Bykov, V.I., Gorban, A.N., \& Elokhin, V.I. (1991). {\it Kinetic 
models of catalytic reactions.} Comprehensive Chemical Kinetics, Vol. 32, 
Compton, R.\ G., ed.  Amsterdam: Elsevier.

\bibitem[Zhu \& Petzold(1999)]{Zhu99}Zhu, W., \&  Petzold, L. (1999).
Model reduction for chemical kinetics: An 
optimization approach. {\it AIChE Journal},  (April 1999), 869-886.

\bibitem[Zmievskii, Kalin \& Deville(2000)]{ZKD00} 
Zmievskii, V.\ B., Karlin, I.\ V., \& Deville, M. (2000). The 
universal limit in dynamics of dilute polymeric solutions. {\it Physica A}, 
{\bf 275(1-2)}, 152-177.





\end{thebibliography}
\end{document}